\documentclass[twocolumn,floatfix,superscriptaddress,a4paper,showkeys,nofootinbib,reprint]{revtex4-1}
\usepackage{epsfig}
\usepackage{latexsym}
\usepackage{xspace}
\usepackage[colorlinks=true,linktocpage=true,linkcolor=blue,citecolor=blue,allcolors=blue]{hyperref}
\usepackage[utf8]{inputenc}
\usepackage{indentfirst}
\usepackage{enumerate}
\usepackage{listings}
\usepackage{color}

\usepackage[caption=false,position=top]{subfig}

\usepackage{amsmath}
\usepackage{amssymb}
\usepackage[english]{babel}
\usepackage{url}

\newcommand{\eq}[1]{\begin{align} #1 \end{align}}
\newcommand{\mean}[1]{\left\langle #1 \right\rangle}

\newcommand{\sNN}{\sqrt{s_{\rm NN}}}

\newcommand{\Bbar}{\bar{B}}
\newcommand{\NB}{N_B}
\newcommand{\NBb}{N_{\bar{B}}}

\newcommand{\NN}{N}
\newcommand{\rb}{r_B}
\newcommand{\chatB}[2]{\hat{c}_{#1,#2}^{B,\bar{B}}}

\begin{document}

\title{Acceptance dependence of factorial cumulants, long-range correlations,\\ and the antiproton puzzle}

\author{Adam Bzdak}
\affiliation{AGH University of Krakow,
Faculty of Physics and Applied Computer Science,
30-059 Krak\'ow, Poland}

\author{Volker Koch}
\affiliation{Nuclear Science Division, Lawrence Berkeley National Laboratory, 1 Cyclotron Road, Berkeley, CA 94720, USA}

\author{Volodymyr~Vovchenko}
\affiliation{Physics Department, University of Houston, 3507 Cullen Blvd, Houston, TX 77204, USA}

\begin{abstract}
We analyze joint factorial cumulants of protons and antiprotons in relativistic heavy-ion collisions and point out that they obey the scaling $\hat{C}_{nm}^{p,\bar{p}} \propto \mean{N_p}^n \mean{N_{\bar{p}}}^m$ as a function of acceptance when only long-range correlations are present in the system, such as global baryon conservation and volume fluctuations.
This hypothesis can be directly tested experimentally without the need for corrections for volume fluctuations.
We show that if correlations among protons and antiprotons are driven by global baryon conservation and volume fluctuations only, the equality $\hat{C}_{2}^{p} / \mean{N_p}^2 = \hat{C}_{2}^{\bar{p}} / \mean{N_{\bar{p}}}^2$ holds for large systems created in central collisions.
We point out that the experimental data of the STAR Collaboration from phase I of RHIC beam energy scan are approximately consistent with the scaling $\hat{C}_{nm}^{p,\bar{p}} \propto \mean{N_p}^n \mean{N_{\bar{p}}}^m$, but the normalized antiproton correlations are stronger than that of protons, $-\hat{C}_{2}^{\bar{p}} / \mean{N_{\bar{p}}}^2 > -\hat{C}_{2}^{p} / \mean{N_p}^2$. 
Existing theoretical baselines, based on global baryon conservation and volume fluctuations, cannot explain the data, to which we refer as the antiproton puzzle.
We also discuss high-order factorial cumulants which can be measured with sufficient precision within phase II of RHIC-BES. 
\end{abstract}


\maketitle

\section{Introduction}
\label{sec:intro}
Exploring the structure of the QCD phase diagram and searching for a possible phase transition and an associated critical point is one of the major goals in strong interaction research (see \cite{Bzdak:2019pkr} for a review). One of the key observables are fluctuations of globally conserved charges in a subsystem of the fireball created in high-energy nucleus-nucleus collisions, most prominently (net) baryon number fluctuations. Fluctuation observables have the advantage that they are accessible both to experiment and theory, such as Lattice QCD. 
However, care needs to be taken to make comparisons between theory and experiment. 
Theoretical calculations typically deal with a system with fixed volume in the grand-canonical ensemble, where the charges are conserved only on the average. 
In addition, they calculate fluctuations of the baryon number. 
In contrast, for the system studied in heavy-ion collision experiments, the charges are globally conserved, and even for the best centrality selection, the volume of the system fluctuates. 
Also, experiments can only measure the fluctuation of protons, and possibly lambdas and light nuclei, but not those of all baryons. 
Therefore, even for a system without any phase transition, the fluctuation observables extracted in the experiment differ from the naive theoretical expectation, and appropriate corrections need to be applied for a meaningful comparison.

In order to establish if experiments see anything interesting that could be possibly related to a phase transition or at least some new dynamics, it is desirable to have a baseline that only contains known (non-critical) physics. 
Such baselines have meanwhile been developed. 
In Ref. \cite{Braun-Munzinger:2020jbk}, the authors use the hadron resonance gas model with the canonical treatment of global baryon number conservation for their baseline. 
The authors of Ref. \cite{Vovchenko:2021kxx} used for their baseline viscous hydrodynamics together with global baryon number conservation. 
In addition, they also considered a repulsive (hard-core) interaction for the baryons, which is constrained to reproduce baryon number susceptibilities from Lattice QCD calculations at vanishing baryon density. 
However, both baselines still make model assumptions, such as the hadron resonance gas or the validity of viscous hydrodynamics, even at the lower energies explored in the RHIC beam energy scan. 
Therefore, it would be desirable to have an experimental test to which extent the observed fluctuations deviate from a scenario where all we have is independent hadron production subject to global baryon number conservation and effects from volume or impact parameter fluctuations.

Here, we discuss such observables, namely the so-called reduced correlation coefficients $\hat{c}_n = \hat{C}_n / \mean{N}^n$, which are the factorial cumulants scaled with proper powers of the mean particle number and have earlier been introduced and discussed in~\cite{Bzdak:2016sxg,Bzdak:2016jxo,Bzdak:2017ltv,Barej:2020ymr}.
We will show that these reduced correlation coefficients should be independent of the acceptance if the observed fluctuations arise solely from a combination of statistical fluctuations, volume fluctuations, and global baryon number conservation. 
We will further show that the second-order reduced correlation coefficient should be (to a very good approximation) the same for protons and antiprotons. 

We then analyze the data of the STAR collaboration from the first phase of the RHIC beam energy scan (BES-I) \cite{STAR:2021iop} to see if the fluctuation data are consistent with this simple baseline or if there are indications for additional, potentially critical dynamics.
Given the statistics limitations of the BES-I data, the analysis here is restricted to second-order coefficients.
Although the BES-I data do not reveal a significant variation in the acceptance dependence, one does observe a remarkably strong splitting in the strength of the second-order reduced correlation coefficient between protons and antiprotons. 
While the protons can be reasonably well described by global baryon conservation-based baselines, the strength of antiproton anticorrelation is significantly underestimated by the available baselines, to which we refer to as the \emph{antiproton puzzle}.
We then discuss the possible mechanisms that may be responsible for the observed splitting, such as the possible incomplete equilibration between stopped and produced matter.

This paper is organized as follows. 
The next section~\ref{sec:defs} introduces the reduced correlation coefficient, where it is shown that they are independent of the acceptance in case of global baryon number conservation as well as volume fluctuations. 
In Sec.~\ref{sec:data}, we show the results based on the STAR BES-I data \cite{STAR:2021iop}, and the discussion of our findings in~\ref{sec:discussion} closes the paper. 

\section{Factorial cumulants and long-range correlations}
\label{sec:defs}

\subsection{Definitions}

Let $P(N_{B},N_{\Bbar})$ be the distribution of two numbers, $N_B$ and $N_{\bar{B}}$, such as the numbers of protons and antiprotons. Then, the moment generating function 
$M_{N_{B},N_{\Bbar}} (t_{B}, t_{\Bbar}) \equiv \mean{e^{t_{B} N_{B}} e^{t_{\Bbar} N_{\Bbar}}}=\sum_{N_{B},N_{\Bbar}}P(N_{B},N_{\Bbar})e^{t_{B} N_{B}} e^{t_{\Bbar} N_{\Bbar}}$ 
defines the moments
\eq{
\mean{N_B^n N_{\bar{B}}^m } = \left. \frac{\partial^{n+m}}{\partial t_B^n \partial t_{\bar{B}}^m} M_{N_B,N_{\bar{B}}} (t_B, t_{\bar{B}}) \right|_{t_B = t_{\bar{B}} = 0},
}
while it's logarithm $G_{N_B,N_{\bar{B}}}(t_B,t_{\bar{B}}) = \ln M_{N_B,N_{\bar{B}}}(t_B,t_{\bar{B}})$ -- the cumulant generating function -- defines the cumulants
\eq{
K_{n,m} =   \left. \frac{\partial^{n+m}}{\partial t_B^n \partial t_{\bar{B}}^m} G_{N_B,N_{\bar{B}}} (t_B, t_{\bar{B}}) \right|_{t_B = t_{\bar{B}} = 0}.
}

The factorial moments read
\eq{
F_{n,m} 
& = \mean{\frac{N_B!}{(N_B-n)!} \frac{N_{\Bbar}!}{(N_{\Bbar}-m)!}} \nonumber \\
& = \left. \frac{\partial^{n+m} }{\partial z_B^n \partial z_{\bar{B}}^m} h_{N_B,N_{\bar{B}}} (z_B, z_{\bar{B}}) \right|_{z_B = z_{\bar{B}} = 1},
}
where  
\eq{
h_{N_B,N_{\bar{B}}} (z_B, z_{\bar{B}}) \equiv \mean{z_B^{N_B} z_{\Bbar}^{N_{\Bbar}}} = M_{N_B,N_{\bar{B}}}(\ln z_B, \ln z_{\Bbar}),
}
is the factorial moment generating function.

Finally, the factorial cumulants, or integrated correlation functions, are defined as
\eq{\label{eq:FCgen}
\hat{C}_{n,m} = \left. \frac{\partial^{n+m} }{\partial z_B^n \partial z_{\bar{B}}^m} \hat{H}_{N_B,N_{\bar{B}}} (z_B, z_{\bar{B}}) \right|_{z_B = z_{\bar{B}} = 1},
}
and they probe genuine~(irreducible) multi-particle correlations~\cite{Bzdak:2016sxg}.
Here $\hat{H}_{N_B,N_{\bar{B}}} (z_B, z_{\bar{B}}) = \ln h_{N_B,N_{\bar{B}}} (z_B, z_{\bar{B}}) = G_{N_B,N_{\bar{B}}} (\ln z_B, \ln z_{\bar{B}})$ is the factorial cumulant generating function.

\subsection{Long-range correlations and binomial acceptance}

Let us denote $N_B^{\rm tot}$ and $N_{\bar{B}}^{\rm tot}$ as the numbers of baryons and antibaryons in full space. For simplicity, let us consider rapidity as the only phase-space variable.
If the correlations in rapidity are long-range only, correlations among the different particles are independent of their rapidities.
In this case, the probability of observing any given particle in a particular rapidity acceptance is independent of all other particles, and the acceptance corrections may be modeled by a binomial probability distribution.
The factorial cumulants $\hat{C}^{B,\bar{B}}_{nm}$ of accepted baryon and antibaryon numbers $N_B$ and $N_{\bar{B}}$ are then given by~\cite{Kitazawa:2011wh,Kitazawa:2012at,Bzdak:2012ab,Savchuk:2019xfg}
\eq{
\hat{C}^{B,\bar{B}}_{nm} = \alpha_B^n \alpha_{\bar{B}}^m \hat{C}^{B_{\rm tot},\bar{B}_{\rm tot}}_{nm},
}
where
\eq{
\alpha_B = \frac{\mean{N_B}}{\mean{N_B^{\rm tot}}}, \qquad \alpha_{\bar{B}} = \frac{\mean{N_{\bar{B}}}}{\mean{N_{\bar{B}}^{\rm tot}}},
}
are the acceptance parameters.
It can be seen that the factorial cumulants of accepted (anti)baryons obey the scaling
\eq{\label{eq:scaling}
\frac{\hat{C}^{B,\bar{B}}_{nm}}{\mean{N_B}^n \mean{N_{\bar{B}}}^m} = \hat{c}^{B,\bar{B}}_{nm} = \rm{const.},
}
where $\hat{c}^{B,\bar{B}}_{nm}$
is the reduced correlation coefficient which is independent of acceptance.
The above considerations also hold for a (uniform) experimental efficiency~\cite{Pruneau:2002yf}.

The experiment measures (anti)protons as a proxy for (anti)baryons.
If all correlations among baryons are isospin blind, one can model this effect as a binomial efficiency correction~\cite{Kitazawa:2012at}, leading the following proton factorial cumulants,
\eq{
\hat{C}^{p,\bar{p}}_{nm} = \alpha_p^n \alpha_{\bar{p}}^m \hat{C}^{B_{\rm tot},\bar{B}_{\rm tot}}_{nm},
}
with $\alpha_{p(\bar{p})} = \mean{{N_p(N_{\bar{p}})}}/\mean{N_{B(\bar{B})}^{\rm tot}}$ and $\mean{N_p(N_{\bar{p}})}$ are the mean numbers of accepted (anti)protons.
Thus, proton factorial cumulants obey the scaling
\eq{\label{eq:scalingprotons}
\frac{\hat{C}^{p,\bar{p}}_{nm}}{\mean{N_p}^n \mean{N_{\bar{p}}}^m} = \hat{c}^{B,\bar{B}}_{nm} = \rm{const.}
}

By testing the relation \eqref{eq:scalingprotons} as a function of rapidity and/or transverse momentum cut, one can probe whether all observed correlations among (anti)protons are long-range only.
Below we analyze two common sources of long-range correlations in heavy-ion collisions.

\subsection{Global baryon conservation}

Let us assume that global baryon conservation is the only source of correlations.
Then, the total numbers of baryons and antibaryons are described by the ideal gas in the canonical ensemble, which corresponds to two Poisson-distributed numbers with a fixed difference corresponding to the conserved baryon number, $B = \NB - \NBb$. The joint probability is proportional to,
\eq{\label{eq:PNBNBar}
P(\NB, \NBb) \propto \frac{\mean{\NB}^{\NB}}{\NB!} \frac{\mean{\NBb}^{\NBb}}{\NBb!} \delta_{B,\NB - \NBb},
}
from which one can derive the following cumulant generating function~\cite{Bzdak:2012an}
\eq{\label{eq:Gcons}
G_{\NB, \NBb}(t_B,t_{\Bbar}) & = \ln \frac{\displaystyle\sum_{\NB,\NBb} P(\NB, \NBb) e^{t_B \NB + t_{\bar{B}} \NBb}}{\displaystyle\sum_{\NB,\NBb} P(\NB, \NBb)} \nonumber \\
& = \ln \left[ \exp\left(B \, \frac{t_B - t_{\bar{B}}}{2}\right) \frac{I_B(2z e^{\frac{t_B + t_{\bar{B}}}{2}})}{I_B(2z)} \right].
}
Here $z = \sqrt{\mean{\NB} \mean{\NBb}}$.

The cumulant generating function~\eqref{eq:Gcons} can be used to calculate the reduced correlation coefficients entering Eq.~\eqref{eq:scaling}.
The resulting expressions are lengthy and not particularly informative.
One can derive more compact expressions by considering the large volume limit, where $B \gg 1$ and/or $z \gg 1$ holds. This holds true for heavy-ion collisions, where the total net baryon number is $B \sim 400$.
We utilize the uniform asymptotic expansion of the modified Bessel function~[see Eq. (9.7.7) in \cite{AbramowitzStegun}]
\eq{\label{eq:IUniAsympt}
I_{\nu} (\nu \tilde{z}) \stackrel{\nu \to \infty}{\sim} \frac{1}{\sqrt{2\pi\nu}} \frac{e^{\nu \eta}}{(1+\tilde{z}^2)^{1/4}} \left\{1 + O(\nu^{-1})\right\},
}
where $\eta = \sqrt{1+\tilde{z}^2} + \ln \frac{\tilde{z}}{1 + \sqrt{1+\tilde{z}^2}}$.

As a result, one can express the cumulant generating function in the following form
\eq{\label{eq:Gasympt}
G_{\NB, \NBb}(t_B,t_{\Bbar}) = \NN f(t_B,t_{\Bbar}) + a(t_B,t_{\Bbar}) + O\left(\NN^{-1} \right).
}
Here $\NN = \mean{\NB} + \mean{\NBb}$.
The first term is the leading-order contribution, i.e. $G_{\NB, \NBb}^{\rm LO}(t_B,t_{\Bbar}) = \NN f(t_B,t_{\Bbar})$, where the function $f(t_B,t_{\Bbar})$ reads
\eq{
f(t_B,t_{\Bbar}) & = \frac{t_B - t_{\Bbar}}{2} \rb - 1 
\nonumber \\
& \quad + 
\sqrt{\rb^2 + (1-\rb^2) e^{t_B + t_{\bar{B}}}}
\nonumber \\
& \quad + 
|\rb| \ln \left[ \frac{(1+|\rb|) e^{\frac{t_B + t_{\bar{B}}}{2}}}{|\rb| + \sqrt{\rb^2 + (1-\rb^2) e^{t_B + t_{\bar{B}}}}} \right],
}
where $\rb = B / \NN$ is the ratio of net-baryon number over the number of baryons plus anti-baryons.
The next-to-leading term, $\delta G_{\NB, \NBb}^{\rm NLO}(t_B,t_{\Bbar}) \equiv a(t_B,t_{\Bbar})$, reads
\eq{
a(t_B,t_{\Bbar}) = [\rb^2 + (1-\rb^2) e^{t_B + t_{\bar{B}}}]^{-1/4}~.
}

Using Eq.~\eqref{eq:FCgen} we evaluate the reduced coefficients $\hat{c}_{n,m}^{B,\bar{B}}$ in Eq.~\eqref{eq:scaling} in the large volume limit. 
The second-order couplings read
\eq{\label{eq:cons2}
\hat{c}_{2,0}^{B,\bar{B}} = -\hat{c}_{1,1}^{B,\bar{B}} = \hat{c}_{0,2}^{B,\bar{B}} = -\frac{1}{\NN}.
}
One can see that baryon conservation generates negative correlations among baryons and among antibaryons and a positive correlation between baryons and antibaryons.
One also sees that the strength of these correlations is identical in the large volume limit.

The third-order couplings read
\eq{\label{eq:cons3}
& \chatB{3}{0} = \frac{3-\rb}{\NN^2}, \quad \chatB{0}{3} = \frac{3+\rb}{\NN^2}, \\
& \chatB{2}{1} = -\frac{1-\rb}{\NN^2}, \quad \chatB{1}{2} = -\frac{1+\rb}{\NN^2}.
}
In contrast to two-particle correlations, here baryon conservation leads to positive three-(anti)baryon correlation, also their strength is no longer equal but depends on baryon/antibaryon ratio.
It also scales with the inverse square of the total number of baryons and antibaryons, $\NN$, thus, the magnitude of the signal is weaker.

Fourth-order couplings read
\eq{\label{eq:cons4}
& \chatB{4}{0} = -\frac{3(5-4\rb+\rb^2)}{\NN^3}, \quad \chatB{0}{4} = -\frac{3(5+4\rb+\rb^2)}{\NN^3}, 
\\
& \chatB{3}{1} = 3\frac{(1-\rb)^2}{\NN^3}, \quad \chatB{1}{2} = 3\frac{(1+\rb)^2}{\NN^3}, 
\\
& \chatB{2}{2} = \frac{1-3\rb^2}{\NN^3} .
}

The scaling given by Eq.~\eqref{eq:scalingprotons} would no longer hold if a source of local correlations is present in addition to global baryon conservation. 
An example of such a case is worked out in the Appendix.

\subsection{Volume fluctuations}

Here, we discuss volume fluctuations, which are unavoidable in heavy-ion collisions, and provide another example of a source of long-range correlations.
We show that if the factorial cumulants obey the scaling~\eqref{eq:scalingprotons} at fixed volume $V$, this scaling is preserved in the presence of volume fluctuations. 
The only assumption here is that all cumulants at fixed volume are linearly proportional to it, such as the case in the model of independent sources~\cite{Gorenstein:2011vq,Holzmann:2024wyd} or thermal systems in thermodynamic limit~\cite{Skokov:2012ds}.

Here, we generalize the result of Ref.~\cite{Skokov:2012ds} to joint factorial cumulants of two quantities.
We assume that the volume $V$ follows a certain probability density $\rho_V(V)$ and its cumulants are characterized by the corresponding cumulant generating function,
\eq{
G_V(t) = \ln \mean{ e^{t\,V}} = \ln \int dV \rho_V(V) e^{t\,V} .
}
The cumulants $\kappa_{n,m} (V)$ at fixed volume $V$ are linearly proportional to the volume $V$, therefore, one can define reduced cumulants
\eq{
\chi_{n,m} = \kappa_{n,m} / V,
}
such that one can introduce cumulant generating function for $\chi_{n,m}$ as
\eq{
G_{N_B,N_{\bar{B}}}^\chi(t_A,t_B) = \frac{G_{N_B,N_{\bar{B}}}(t_A,t_B)}{V}.
}

The moment generating function $M_{\tilde N_B,\tilde N_{\bar{B}}}$
is obtained by folding the moment generating function at fixed volume, $M_{N_B,N_{\tilde{B}}} = \mean{e^{t_{B} N_{B}+t_{\Bbar} N_{\Bbar}}}_V=\sum_{N_{B},N_{\Bbar}}P_V(N_{B},N_{\Bbar})e^{t_{B} N_{B}+ t_{\Bbar} N_{\Bbar}}$ with the probability distribution $\rho_V(V)$
\begin{align}
   M_{\tilde N_B,\tilde N_{\bar{B}}} =\int dV \rho_V(V) \mean{e^{t_{B} N_{B}+t_{\Bbar} N_{\Bbar}}}_V
\end{align}
Here $P_V(N_{B},N_{\Bbar})$ is the baryon and anti-baryon distribution at fixed volume and  $\mean{\ldots}_V$ indicates averaging at a fixed volume. We thus obtain the cumulant generating function $G_{\tilde N_B,\tilde N_{\bar{B}}}$ for cumulants $\tilde{\kappa}_{n,m}$ affected by volume fluctuations 
\eq{
G_{\tilde N_B,\tilde N_{\bar{B}}}(t_A,t_B) & = \ln \mean{e^{t_B \tilde N_B + t_{\bar{B}} \tilde N_{\bar{B}}}}
\nonumber \\
& = \ln \int dV \rho_V(V) \displaystyle\mean{e^{t_B N_B + t_{\bar{B}} N_{\bar{B}}}}_V.
}
Given that, by definition, $G_{N_B,N_{\bar{B}}}(t_B,t_{\bar{B}}) = \ln \mean{e^{t_B N_B + t_{\bar{B}} N_{\bar{B}}}}_V$, one has
\eq{
\mean{e^{t_B N_B + t_{\bar{B}} N_{\bar{B}}}}_V = e^{G_{N_B,N_{\bar{B}}}(t_B,t_{\bar{B}})} = e^{V G_{N_B,N_{\bar{B}}}^\chi(t_B,t_{\bar{B}})},
}
therefore,
\eq{
G_{\tilde N_B,\tilde N_{\bar{B}}}(t_B,t_{\bar{B}}) & = \ln \int dV \rho_V(V) e^{V G_{N_B,N_{\bar{B}}}^\chi(t_B,t_{\bar{B}})}
\nonumber \\
& = \ln \mean{e^{V G_{N_B,N_{\bar{B}}}^\chi(t_B,t_{\bar{B}})}}
\nonumber \\
& = G_V[G_{N_B,N_{\bar{B}}}^\chi(t_B,t_{\bar{B}})].
}

The factorial cumulant generating function, therefore, reads
\eq{
\hat{H}_{\tilde N_B,\tilde N_{\bar{B}}} (z_B, z_{\bar{B}}) 
& \equiv G_{\tilde N_B,\tilde N_{\bar{B}}} (\ln z_B, \ln z_{\bar{B}})
\nonumber \\
& = G_V[H_{N_B,N_{\bar{B}}}^\chi (z_B, z_{\bar{B}})]. 
}
Here $\hat{H}_{N_B,N_{\bar{B}}}^\chi (z_B, z_{\bar{B}}) = G_{N_B,N_{\bar{B}}}^\chi(\ln t_B,\ln t_{\bar{B}})$.
The factorial cumulants $\hat{\tilde{C}}_{n,m}$ subject to volume fluctuations are then computed as
\eq{
\hat{\tilde{C}}_{n,m} = \left. \frac{\partial^{n+m} }{\partial z_B^n \partial z_{\bar{B}}^m} G_V[\hat{H}_{N_B,N_{\bar{B}}}^\chi (z_B, z_{\bar{B}})] \right|_{z_B = z_{\bar{B}} = 1},
}
which can be evaluated using the combinatorial form of Fa\'a di Bruno's formula for multivariate derivatives of composite functions~\cite{haiman1989incidence,Vovchenko:2021yen}
\eq{
\hat{\tilde{C}}_{n,m} = \sum_{\pi \in \Pi_{n+m}} \kappa_{|\pi|} [V] \, \prod_{b \in \pi} \hat{C}_{b_n,|b|-b_n}^{\chi}. 
}
Here $\Pi_{n+m}$ is the set of all partitions of the set $\{1 \ldots (n+m) \}$ into blocks, $\pi$ is a single partition from $\Pi_{n+m}$, 
$|\pi|$ is the number of blocks in partition $\pi$,
$b$ is a single block from $\pi$,
$b_n$ is the number of elements in the block $b$ that are smaller or equal than $n$, $|b|$ is the total number of elements in the block $b$,
and
\eq{
\hat{C}_{b_n,|b|-b_n}^{\chi} = \left. \frac{\partial^{|b|} }{\partial z_B^{b_n} \partial z_{\bar{B}}^{|b|-b_n}} \hat{H}_{N_B,N_{\bar{B}}}^\chi (z_B, z_{\bar{B}}) \right|_{z_B = z_{\bar{B}} = 1}.
}
Given that 
\eq{
\hat{C}_{b_n,|b|-b_n}^{\chi} 
& = \frac{\hat{C}_{b_n,|b|-b_n}}{\mean{V}}
\nonumber \\ 
& = \mean{N_A}^{b_n} \mean{N_B}^{|b|-b_n} \frac{\hat{c}_{b_n,|b|-b_n}}{\mean{V}},
}
and $\sum_B b_n = n$, one obtains,
\eq{
\hat{\tilde{C}}_{n,m} & = \mean{N_B}^{n} \mean{N_{\bar{B}}}^{m} \sum_{\pi \in \Pi_{n+m}} \frac{\kappa_{|\pi|} [V]}{\mean{V}^{|\pi|}} \, \prod_{b \in \pi} \hat{c}_{b_n,|b|-b_n} 
\nonumber \\
& = \mean{N_B}^{n} \mean{N_{\bar{B}}}^{m} \hat{\tilde{c}}_{n,m},
}
where we introduce the coupling $\hat{\tilde{c}}_{n,m}$ in the presence of volume fluctuations,
\eq{
\hat{\tilde{c}}_{n,m} = \sum_{\pi \in \Pi_{n+m}} \frac{\kappa_{|\pi|} [V]}{\mean{V}^{|\pi|}} \, \prod_{b \in \pi} \hat{c}_{b_n,|b|-b_n}. 
}

One can therefore see that the presence of volume fluctuations preserves the scaling $\hat{\tilde{C}}_{n,m} \propto \mean{N_B}^{n} \mean{N_{\bar{B}}}^{m}$ as long as this scaling was present at a fixed volume, i.e. as long as $\hat{c}_{b_n,|b|-b_n}$ are independent of $\mean{N_B}$ and $\mean{N_{\bar{B}}}$.
For example, one has
\eq{
\label{eq:volF}
\hat{\tilde{c}}_{i,j} & = \hat{c}_{i,j} + \frac{\kappa_2[V]}{\mean{V}^2}, \quad \text{for} \quad i+j = 2.
}
This quantity is independent of the acceptance, and the effect of volume fluctuations is a constant shift by $\frac{\kappa_2[V]}{\mean{V}^2}$.
One can construct combinations of second-order couplings that are unaffected by volume fluctuations, such as
\eq{
\label{eq:SIQ1}
\hat{\tilde{c}}_{2,0} - \hat{\tilde{c}}_{0,2} & = \hat{c}_{2,0} - \hat{c}_{0,2}, \\
\label{eq:SIQ2}
\hat{\tilde{c}}_{2,0} + \hat{\tilde{c}}_{0,2} - 2 \hat{\tilde{c}}_{1,1} & = \hat{c}_{2,0} + \hat{c}_{0,2} -  2 \hat{c}_{1,1}~.
}
These quantities can be connected to the strongly intensive measures $\Delta$ and $\Sigma$ discussed in Refs.~\cite{Gorenstein:2011vq,Broniowski:2017tjq}, as well as strongly intensive cumulants of Ref.~\cite{Sangaline:2015bma}.

\section{Analysis of experimental data}
\label{sec:data}

\subsection{BES-I data}

\begin{figure*}[t]
    \includegraphics[width=.99\textwidth]{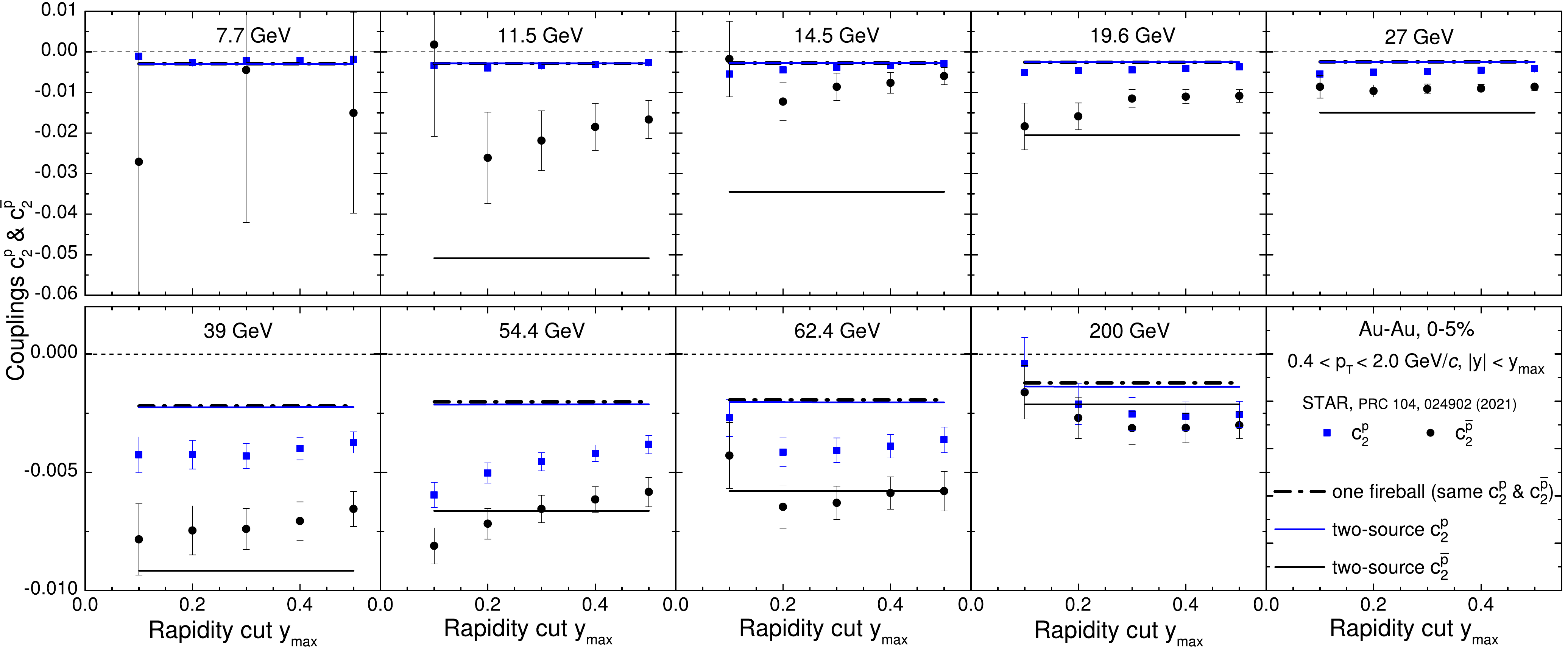}
    \caption{
    Rapidity acceptance dependence of the second-order normalized factorial cumulant $\hat{C}_2/(\hat{C}_1)^2$ for protons~(blue symbols) and antiprotons~(black symbols), as measured by the STAR Collaboration within RHIC-BES-I~\cite{STAR:2020tga,STAR:2021iop}.
    The dashed line shows the results~(identical for protons and antiprotons) from a single fireball model incorporating exact global baryon conservation~\cite{Vovchenko:2021kxx}.
    The solid lines depict the predictions of the two-source model.
    }
    \label{fig:STARC2C1scall}
\end{figure*}

The STAR Collaboration has presented measurements of ordinary and factorial cumulants for protons and antiprotons in Au-Au collisions at $\sNN = 7.7-200$~GeV in~\cite{STAR:2020tga,STAR:2021iop} within phase I of RHIC beam energy scan.
The absolute values of the cumulants as well as the scaled factorial cumulants such as  $\hat{C}_n^{p} / \mean{N_p}$ or $\hat{C}_n^{{\bar{p}}} / \mean{N_{\bar{p}}}$  were published in \cite{STAR:2021iop}, but not the corresponding couplings $\hat{c}_n^{p} = \hat{C}_n^{p} / \mean{N_p}^n$ etc.
Figure~\ref{fig:STARC2C1scall} shows the rapidity cut dependence of the couplings for protons, $\hat{C}_2^p / \mean{N_p}^2$~(blue), and antiprotons, $\hat{C}_2^{{\bar{p}}} / \mean{N_{\bar{p}}}^2$~(black).
The experimental data indicate that the scalings $\hat{C}_{2}^{p} / \mean{N_p}^2 \propto \rm const$ and $\hat{C}_{2}^{{\bar{p}}} / \mean{N_{\bar{p}}}^2 \propto \rm const$ approximately hold, with some indications for additional negative two-proton correlations at small $y_{\rm cut}$ at $\sNN = 14.5-27$~GeV.
Interestingly, one can clearly see that anticorrelations of antiprotons are notably stronger than those of the protons at energies below $\sNN \leq 62.4$~GeV, where $-\hat{C}_{2}^{{\bar{p}}} / \mean{N_{\bar{p}}}^2 > -\hat{C}_{2}^{p} / \mean{N_p}^2$.

\subsection{Single fireball model}

We now analyze the data within phenomenological models.
In a single fireball model, 
for example, single-fluid hydrodynamics,
all baryons and antibaryons are emitted from a single thermalized source.
In the absence of any interactions among baryons and neglecting volume fluctuations, the joint probability is given by Eq.~\eqref{eq:PNBNBar} where $B = N_{\rm part}$.
The couplings are given by Eq.~\eqref{eq:cons2} where $N = \mean{N_B} + \mean{N_{\bar{B}}}$.
One can express the mean numbers of (anti)baryons as
\eq{
\mean{B} = N_{\rm part} + \mean{N_{\rm pairs}^{B\bar{B}}}, \quad \mean{\bar{B}} = \mean{N_{\rm pairs}^{B\bar{B}}},
}
where $\mean{N_{\rm pairs}^{B\bar{B}}}$ is the mean (net) number of produced $B\bar{B}$ pairs.
The couplings read
\eq{\label{eq:c2singlecomp}
\hat{c}_2^p = \hat{c}_2^{\bar{p}} = -\hat{c}_{1,1}^{p\bar{p}} = -\frac{1}{N_{\rm part} + 2\mean{N_{\rm pairs}^{B\bar{B}}}}.
}

To make model predictions, we need reliable estimates of $N_{\rm part}$, and, especially $\mean{N_{\rm pairs}^{B\bar{B}}}$.
Here we utilize the results for $\mean{B}$ and $\mean{\bar{B}}$ from hydrodynamic simulations within MUSIC~\cite{Shen:2020jwv,Vovchenko:2021kxx}.
The resulting values at different energies are listed in Table~\ref{tab:twocompMUSIC}. 
Of course, since $N_{\rm part}$ is estimated experimentally there is, in principle, no need to invoke additional model calculations for this quantity.
However, we keep the values of $N_{\rm part}$ from MUSIC for consistency, and,  as shown in Table~\ref{tab:twocompMUSIC}, these values are in good agreement with experimental estimates.

\begin{table}
    \centering
    \begin{tabular}{c|c|c}
        $\sNN$~[GeV] & $N_{\rm part}$ & $\mean{N_{\rm pairs}^{B\bar{B}}}$ \\
    \hline
        7.7 & 333~(337 $\pm$ 2) & 3.93 \\
        11.5 & 336*~(338 $\pm$ 2) & 9.83*  \\
        14.5 & 339~(340 $\pm$ 2) & 14.5  \\
        19.6 & 341~(338 $\pm$ 2) & 24.4  \\
        27 & 344~(343 $\pm$ 2) & 33.4  \\
        39 & 343~(342 $\pm$ 2) & 54.6 \\
        54.4 & 342*~(346 $\pm$ 2)  & 75.4* \\
        62.4 & 341~(347 $\pm$ 3) & 86.2  \\
        200 & 346~(351 $\pm$ 2) & 235 \\
    \end{tabular}
    \caption{
    Parameters of the two-component model taken from hydrodynamics simulations within MUSIC~\cite{Shen:2020jwv,Vovchenko:2021kxx}.
    The brackets in the second column correspond to experimental data for $N_{\rm part}$~\cite{STAR:2021iop}.
    There is no MUSIC simulations available at energies 11.5 and 54.4~GeV, the parameters there were obtained through linear interpolation~(marked by *).
    }
    \label{tab:twocompMUSIC}
\end{table}

As follows from Eq.~\eqref{eq:c2singlecomp}, the single-fireball model predicts identical couplings for protons and antiprotons, the numerical values are shown in Fig.~\ref{fig:STARC2C1scall} by the dash-dotted black lines~(for $\sNN \lesssim 54.4$~GeV these coincide with the blue lines).
One can see that the model describes relatively well the data for protons, with the underestimation of the data visible at $\sNN \gtrsim 27$~GeV.
The observed antiproton anticorrelations are significantly underestimated at all energies except $\sNN = 200$~GeV.
The model predictions are essentially the same as those of the non-critical hydrodynamics baseline of Ref.~\cite{Vovchenko:2021kxx} incorporating baryon conservation only.

\begin{figure}[t]
    \includegraphics[width=.5\textwidth]{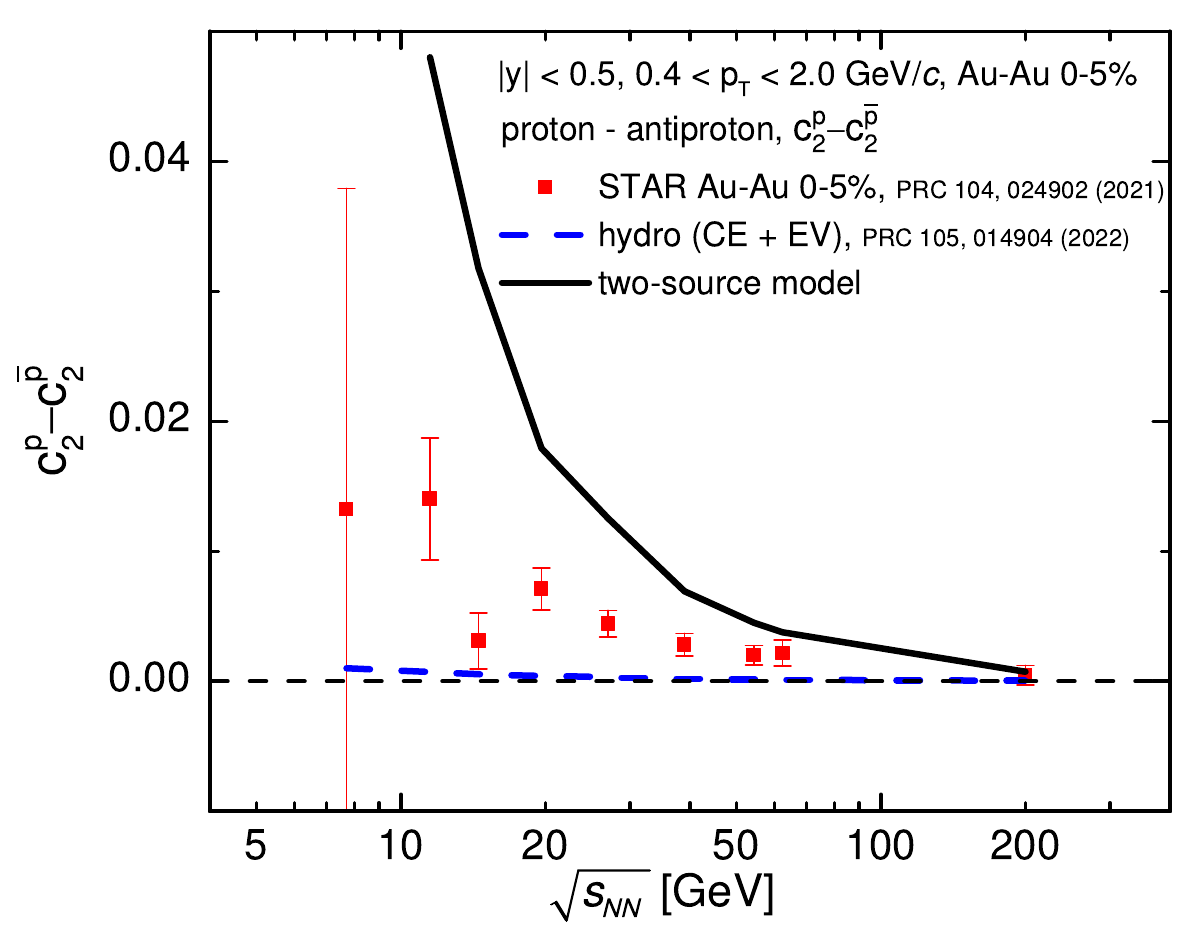}
    \caption{
    Collision energy dependence of the difference $\hat{c}_2^p - \hat{c}_2^{\bar{p}}$ between proton and antiproton couplings for rapidity cut $|y| < 0.5$, as measured by the STAR Collaboration within RHIC-BES-I~(symbols)~\cite{STAR:2020tga,STAR:2021iop}.
    The dashed line corresponds to the vanishing splitting, which corresponds to any model where the only source of fluctuations are global baryon conservation and volume fluctuations.
    The dashed blue line shows the result from hydrodynamic simulations with baryon conservation and excluded volume effect~\cite{Vovchenko:2021kxx}. 
    The solid black line corresponds to the two-component model.
    }
    \label{fig:STARdelta}
\end{figure}

The splitting between $c_2^p$ and $c_2^{\bar{p}}$ is shown in Fig.~\ref{fig:STARdelta} as a function of collision energy.
The single-fireball model, and, in fact, any model where global baryon conservation and volume fluctuations are the only sources of correlations, predicts $c_2^p = c_2^{\bar{p}}$ while the data clearly show non-zero splitting.
If one incorporates the additional repulsion via excluded volume, one could improve the description of protons at $\sNN \gtrsim 27$~GeV, however, antiprotons are still significantly underestimated, as was shown in Ref.~\cite{Vovchenko:2021kxx} for factorial cumulants ratios $\hat{C}_2^{\bar{p}}/\hat{C}_1^{\bar{p}}$.
Excluded volume does generate a small splitting between $c_2^p$ and $c_2^{\bar{p}}$~(dashed blue line in Fig.~\ref{fig:STARdelta}) and suppresses $c_2^{p(\bar{p})}$ at small $y_{\rm max}$, but not nearly enough to describe the sizable difference between protons and antiprotons seen in the data.

Thus, baryon conservation alone is insufficient to describe neither protons nor antiprotons at $\sNN \gtrsim 14.5$~GeV.
Adding volume fluctuations would not help as it will make two-particle correlations more positive~[Eq.~\eqref{eq:volF}] rather than more negative as indicated by the data, and will not generate any splitting between $\hat{c}_2^p$ and $\hat{c}_2^{\bar{p}}$. 

Thus, the description of antiproton cumulants within a hydrodynamics-based single fireball picture remains an open challenge -- the antiproton puzzle -- that needs further investigation.
One possible avenue is the incorporation of re-scattering in the hadronic phase and baryon annihilation, the latter are expected to affect  antiprotons significantly~\cite{Becattini:2016xct} and leave an imprint on their cumulants~\cite{Savchuk:2021aog}.

\subsection{Two-source model}

To address the difference between protons and antiprotons, we consider here a simple two-component model.
It is based on the observation that the measured protons contain both stopped and produced particles, while antiprotons correspond to produced particles only.
We can now formulate a two-source model, where the first source is the stopped matter governed by stopping fluctuations while the second source is the produced matter that is net-baryon free.\footnote{Similar variations of this model for net-proton cumulants were recently considered in Refs.~\cite{Pruneau:2019baa,Savchuk:2024ykb}.}
Neglecting the correlations between the two sources, the antiprotons receive contributions from the second source only.
Assuming that produced matter thermalizes, and following Eqs.~\eqref{eq:scalingprotons} and \eqref{eq:cons2} and neglecting volume~(participant) fluctuations, one obtains
\eq{
\hat{C}^{\bar{p}}_2 = -\frac{\mean{N_{\bar{p}}}^2}{2 \mean{N_{\rm pairs}^{B\bar{B}}}},
}
where $\mean{N_{\rm pairs}^{B\bar{B}}} = (\mean{N_B}_{\rm prod} + \mean{N_{\bar{B}}}_{\rm prod})/2 = \mean{N_{\bar{B}}}$ is the mean number of $B\bar{B}$ pairs produced in the collision.

The protons contain contributions from both stopped and produced matter, $\mean{N_p} = \mean{N_p}_{\rm stopped} + \mean{N_p}_{\rm prod}$.
Since the produced matter is net-baryon free, one has $\mean{N_p}_{\rm prod} = \mean{N_{\bar{p}}}$ and thus $\mean{N_p}_{\rm stopped} = \mean{N_p} - \mean{N_{\bar{p}}}$.
Here we model the stopped protons by a binomial distribution with Bernoulli probability of $p=\mean{N_p}_{\rm stopped}/N_{\rm part}$.
Since the two sources are assumed to be independent, the cumulants simply add. Thus, assuming a fixed number of participants, the second-order proton factorial cumulant reads
\eq{
\hat{C}^{p}_2 = -\frac{(\mean{N_p} - \mean{N_{\bar{p}}})^2}{N_{\rm part}} - \frac{\mean{N_{\bar{p}}}^2}{2 \mean{N_{\rm pairs}^{B\bar{B}}}}.
}
The scaled factorial cumulants read
\eq{
\hat{c}_2^p & = \frac{\hat{C}^{p}_2}{\mean{N_p}^2} = -\frac{(1-R_{\bar{p}p})^2}{N_{\rm part}} - \frac{R_{\bar{p}p}^2}{2 \mean{N_{\rm pairs}^{B\bar{B}}}}, 
\label{eq:twocomp}
\\
\hat{c}_2^{\bar{p}} & = \frac{\hat{C}^{\bar{p}}_2}{\mean{N_{\bar{p}}}^2} = -\frac{1}{2 \mean{N_{\rm pairs}^{B\bar{B}}}}.
\label{eq:twocomppbar}
}
Here $R_{\bar{p}p} = \mean{N_{\bar{p}}}/\mean{N_p}$ is the antiproton-to-proton ratio.
Given that $\mean{N_p}_{\rm prod} = \mean{N_{\bar{p}}}$, this ratio
indicates the fraction of measured protons that are the produced.
One can see that the reduced factorial cumulant for antiprotons is independent of acceptance, while for protons this statement holds only as long as $R_{\bar{p}p}$ is independent of acceptance as well.

One can also calculate the covariance.
Since the two sources are assumed to be independent, and there are no stopped antiprotons, the correlations exist only between produced protons and antiprotons,
\eq{
\hat{C}^{p\bar{p}}_{1,1} & = \frac{\mean{N_{\bar{p}}}^2}{2 \mean{N_{\rm pairs}^{B\bar{B}}}}, \\
\hat{c}^{p\bar{p}}_{1,1} & = \frac{R_{\bar{p}p}}{2 \mean{N_{\rm pairs}^{B\bar{B}}}}.
}
As in the case of proton $\hat{c}_2^{\bar{p}}$, the normalized covariance $\hat{c}^{p\bar{p}}_{1,1}$ is independent of acceptance as long as $R_{\bar{p}p}$ is.

The results for $\hat{c}_2^{p}$ and $\hat{c}_2^{\bar{p}}$ are shown in Fig.~\ref{fig:STARC2C1scall} by solid blue and black lines, respectively.
The results for protons are almost identical to the single-fireball model, while those for antiprotons are shifted down considerably.
This can be understood as follows: the antiprotons now originate only from the produced matter which has a smaller conservation volume as compared to the single fireball case. This results in larger canonical corrections. 

The two-source model thus qualitatively describes the splitting between protons and antiprotons.
However, as one can see from Fig.~\ref{fig:STARC2C1scall}, this splitting is overestimated at energies $\sNN \leq 39$~GeV.
This may indicate that the two sources are not completely independent but also not fully thermalized, as the latter scenario would correspond to the one-component model where there is no splitting.
It should be noted that produced matter is assumed to be thermalized here, such that the cumulants of produced protons and antiprotons are described by the ideal gas in the canonical ensemble.
In the opposite case~\cite{Tomasik:2019ejy}, one may assume that produced $B\bar{B}$ pairs are uncorrelated, and their distribution follows the Poisson statistics.
In this case, factorial cumulants of antiprotons are zero starting from the second order, in contrast to the data.
The experimental data, which lie in-between these two scenarios at $\sNN \lesssim 39$~GeV, may thus suggest an incomplete equilibration of produced matter.

We note in the case of a two-source model, volume fluctuations become more tricky, as here, one may, in principle, have to consider two volumes, one for the participant system and one for the produced system.
For instance, if participant fluctuations are negligible compared to fluctuations of the volume for the produced matter, i.e. of $N_{\rm pairs}^{B\bar{B}}$, one will expect volume fluctuations to affect antiproton $\hat{c}_2^{\bar{p}}$ more strongly, and such mechanism could help in describing the experimental data.

\subsection{Short-range repulsion}

\begin{figure}[t]
    \includegraphics[width=.5\textwidth]{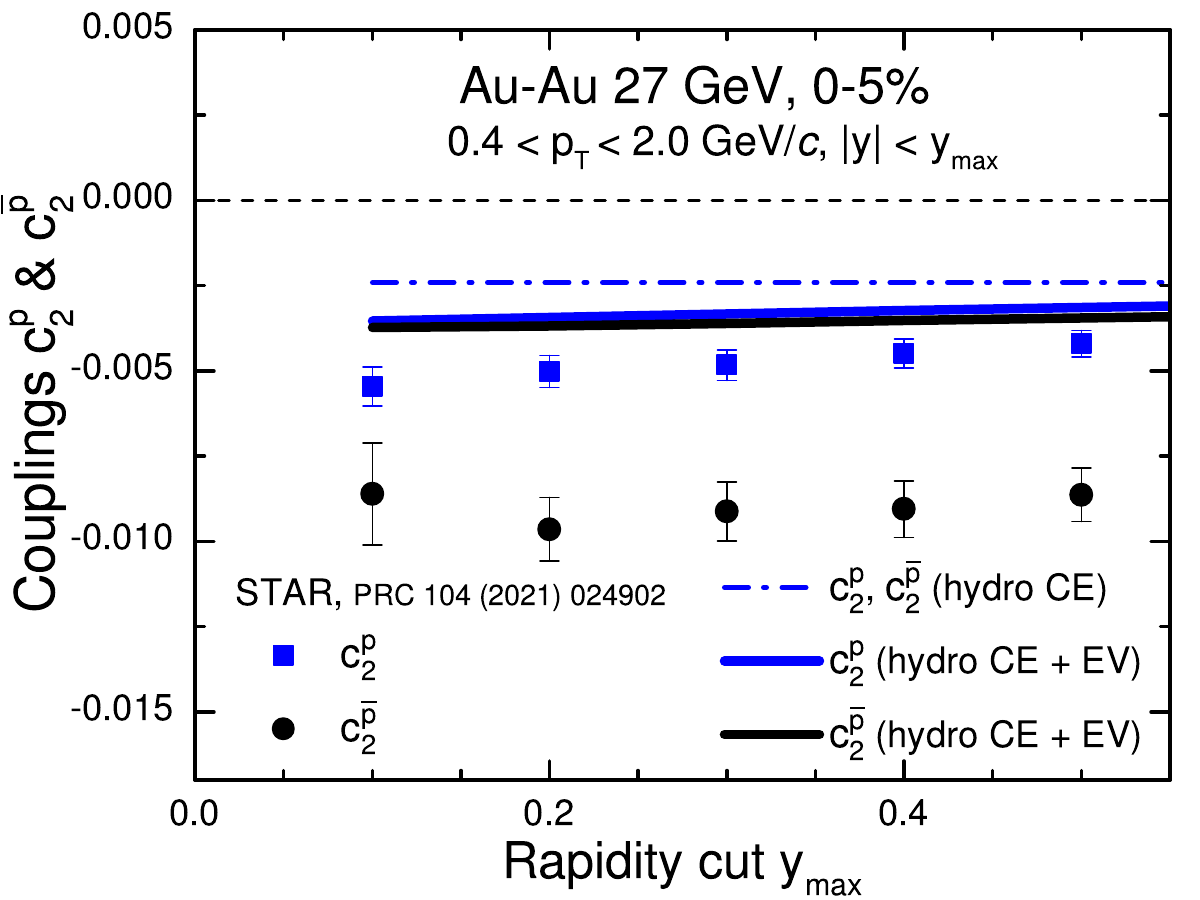}
    \caption{
    Rapidity acceptance dependence of the second-order normalized factorial cumulant $\hat{C}_2/(\hat{C}_1)^2$ for protons~(blue symbols) and antiprotons~(black symbols) in central Au-Au collisions at $\sNN = 27$ GeV, as measured by the STAR Collaboration within RHIC-BES-I~\cite{STAR:2020tga,STAR:2021iop}~(symbols) and computed within non-critical hydrodynamics with global baryon conservation and baryon excluded volume~\cite{Vovchenko:2021kxx}.
    The dashed line shows the results~(identical for protons and antiprotons) that do not incorporate baryon excluded volume~\cite{Vovchenko:2021kxx}.
    }
    \label{fig:STARhydro27}
\end{figure}

The non-critical hydrodynamics baseline of Ref.~\cite{Vovchenko:2021kxx} incorporates the effect of short-range repulsion among baryons through excluded volume.
This leads to the suppression of baryon fluctuations similar to what is observed in lattice QCD and should in principle break the scaling $C_2^p / (C_1^p)^2 = \rm const$ as a function of acceptance.
Figure.~\ref{fig:STARhydro27} depicts the results from the hydrodynamics model of Ref.~\cite{Vovchenko:2021kxx} for $c_2^p$ and $c_2^{\bar{p}}$ for $\sNN = 27$~GeV~(the results for other collision energies are qualitatively the same). 
The solid lines include the excluded volume effect while the dash-dotted line neglects it.
One can see that excluded volume leads to the emergence of a very weak $y_{\rm max}$ dependence. 
Rather, the main effect of the short-range repulsion is the overall shift of the curve by (almost) constant factor.

These results indicate that non-critical correlations such as excluded volume would be very challenging to identify through the analysis of $c_2^p = \rm const$ scaling and its possible violation, at least with second order cumulants.
On the other hand, critical fluctuations are expected to yield a more pronounced acceptance dependence of factorial cumulants~\cite{Ling:2015yau}, especially beyond the second order.

\subsection{LHC energies}

The ALICE Collaboration has performed measurements of (net) proton number cumulants in Pb-Pb collisions at $\sNN = 2.76$~TeV~\cite{ALICE:2019nbs} and $\sNN = 5.02$~TeV~\cite{ALICE:2022xpf}.
The main focus has been on the variance of net proton number normalized by the Skellam (Poisson statistics) baseline, $R_1 = \frac{\kappa_2(N_p-N_{\bar{p}})}{\mean{N_p + N_{\bar{p}}}}$~\cite{ALICE:2019nbs}.
In terms of factorial cumulants, this quantity reads
\eq{
R_1 & = \frac{\kappa_2(N_p-N_{\bar{p}})}{\mean{N_p + N_{\bar{p}}}} 
\nonumber \\
& = 1 + \frac{\hat{C}^{p\bar{p}}_{2,0} + \hat{C}^{p\bar{p}}_{0,2}  - 2\hat{C}^{p\bar{p}}_{1,1} }
{\mean{N_p + N_{\bar{p}}}}.
}
Small deviations from unity are observed in the data, which have been interpreted as being driven by global~\cite{ALICE:2019nbs,Vovchenko:2020kwg} or local~\cite{Savchuk:2021aog,Braun-Munzinger:2023gsd,Vovchenko:2024pvk} baryon number conservation.

Here we advocate for the measurements of the quantity
\eq{
r_1 & \equiv \frac{R_1 - 1}{\mean{N_p + N_{\bar{p}}}} \nonumber \\
&= \frac{\hat{C}^{p\bar{p}}_{2,0} + \hat{C}^{p\bar{p}}_{0,2}  - 2\hat{C}^{p\bar{p}}_{1,1} }
{\mean{N_p + N_{\bar{p}}}^2}
}
as a function of acceptance.
At LHC energies the mean numbers of protons and antiprotons are approximately equal, $\mean{N_p} \approx \mean{N_{\bar{p}}}$, therefore,
$\mean{N_p+N_{\bar{p}}} \approx 2 \mean{N_p} \approx 2 \mean{N_{\bar{p}}}$ and
\eq{\label{eq:r1}
r_1
\approx \frac{1}{4} 
\left( \hat{c}^{p\bar{p}}_{2,0} + \hat{c}^{p\bar{p}}_{0,2}  - 2\hat{c}^{p\bar{p}}_{1,1} \right).
}
As follows from Eq.~\eqref{eq:SIQ2}, $r_1$ is unaffected by volume fluctuations.
If global baryon conservation is the only source of proton correlations,
this quantity is independent of acceptance and, following Eq.~\eqref{eq:cons2}, equals
\eq{
r_1 \approx \frac{1}{\mean{N_B^{\rm tot} + N_{\bar{B}}^{\rm tot}}}.
}

The ALICE Collaboration has presented pseudorapidity acceptance dependence of $R_1$ in~\cite{ALICE:2019nbs,ALICE:2022xpf}, but unfortunately not that of $\mean{N_p + N_{\bar{p}}}$.
This makes it impossible to verify the scaling of $r_1$ with acceptance using available published data.
In~\cite{ALICE:2019nbs} the values of $\mean{N_p + N_{\bar{p}}}$ for the largest acceptance $|\eta| < 0.8$ have been published, for $\sNN = 2.76$~TeV Pb-Pb collisions.
With $R_1 = 0.971 \pm 0.015$ and $\mean{N_p + N_{\bar{p}}} = 36.49 \pm 0.65$, one can estimate $r_1 = (-0.77 \pm 0.40) \cdot 10^{-3}$ and thus
\eq{
\mean{N_B^{\rm tot} + N_{\bar{B}}^{\rm tot}}|_{2.76~\rm TeV} = 1290 \pm 667
}
for the mean total number of baryons and antibaryons in the conservation volume around midrapidity in 0-5\% central Pb-Pb collisions at $\sNN = 2.76$~TeV, if $r_1$ is driven by baryon conservation only.
We advocate for the further analysis of the available experimental data to study the acceptance dependence of $r_1$ in Pb-Pb collisions at both $\sNN = 2.76$ and $5.02$~TeV and testing the possible the scaling $r_1 \propto \rm const$.
Of course, apart from the $r_1$ quantity discussed here, measurements of all the other scaled factorial cumulants given by Eq.~\eqref{eq:scalingprotons} would be of interest as well.

\section{Discussion and summary}
\label{sec:discussion}

In this paper we have shown that the reduced correlation coefficients $\hat{c}_{i,j}$ are independent of the acceptance (in rapidity) if the observed fluctuations arise only from that of an ideal (Poissonian) system
subject to global charge conservation. We further showed that volume fluctuation does not change this behavior. We then extracted these coefficients from the STAR BES-I data \cite{STAR:2021iop} and found the following:
\begin{itemize}
    \item Except for $\sNN = 54.4 \,\rm GeV$, the STAR data from BES-I for both protons and anti-protons are largely consistent with the predicted flat behavior as a function of acceptance. 
    \item However, the data show a clear difference between the reduced correlators of protons and anti-protons, except for the highest energy of $\sNN = 200$~GeV.
   In particular the reduced correlator of anti-protons  is not described by the non-critical hydrodynamics baseline or in other words we do have an \emph{antiproton puzzle.} 
    Here we considered the extreme scenario of a two source model with separate sources for produced and stopped protons. While this model gives the observed trend it clearly over-predicts the effect. We have not considered proton -- anti-proton annihilation which at the lowest energies would lead to different results for protons and anti-protons. However, we note that the difference is still present at a beam energy of $\sqrt{s}=62.4 \,\rm GeV$ where annihilation effects should be similar for protons and anti-protons. 
    \item We have shown that the effect of volume fluctuations also leads to flat acceptance dependence of the reduced correlators. 
    Therefore, it would be interesting if the STAR data without the so-called Centrality Bin Width Correction (CBWC) \cite{Luo:2013bmi} also exhibited a flat behavior. 
    If not, this implies that the CBWC procedure does more than just suppress volume fluctuations.  
    \item Fluctuations of proton and anti-proton numbers have also been measured at the LHC by the ALICE Collaboration \cite{ALICE:2019nbs}. 
    Unfortunately, so far, ALICE has only published the acceptance dependence of net-protons, and it would be interesting to see if the reduced correlators for protons and anti-protons, such as Eq.~\eqref{eq:r1}, are again independent of acceptance. 
    An analysis can also be performed for high-order factorial cumulants, and possible deviations from a flat acceptance dependence can probe the existence of short-range correlations, such as local baryon conservation or chiral criticality.
    \item While our two-source model quantitatively over-predicts the difference between the reduced correlators between protons and anti-protons, one could argue that the STAR data suggest that we are dealing with more than one single source. However, before such a conclusion can be reached, the effect of baryon annihilation needs to be studied in detail.
    Additionally, the effect of local baryon conservation at non-zero baryon density needs to be considered. 
    Measurements of balance functions, which can be connected to the two-particle correlation coefficients studied here may be particularly useful for clarifying the situation~\cite{Bass:2000az,Pruneau:2022mui}.
    
\end{itemize}

To summarize, if the systems created in heavy ion collisions exhibit only long range correlations, the reduced correlators should be independent of the acceptance. This would, for example, be the case for correlations induced by global baryon number conservation and volume fluctuations but could also be due to some other dynamics. 
In the presence of short-range correlations, the reduced correlators will no longer be independent of acceptance.
However, the effects may be small and not visible in the BES-I data from STAR, as can be seen in Fig.~\ref{fig:STARhydro27}, where we show the coupling with and without excluded volume corrections. 
Furthermore, the effects of the possible critical point are expected to be more prominent in high-order (non-Gaussian) correlators. 
Therefore, we are looking forward to an analysis of the reduced correlators from the high statistics BES-II data.

To establish quantitative critical point expectations, one can analyze the behavior of reduced correlators in a microscopic simulation framework, such as molecular dynamics with a critical point~\cite{Kuznietsov:2022pcn,Kuznietsov:2024xyn}.
Apart from critical fluctuations, the acceptance dependence of reduced correlators could also be used to analyze short-range correlations induced by local baryon conservation~\cite{Vovchenko:2024pvk} or baryon annihilation~\cite{Savchuk:2021aog}.
Baryon annihilation, in particular, is expected to induce correlations among protons and antiprotons at small relative momenta, thus, the analysis of $c_{1,1}^{p\bar{p}}$ might be useful to constrain this effect.
These extensions will be the subjects of future studies.

\begin{acknowledgments}
\emph{Acknowledgments.} 
V.K. has been supported by the U.S. Department of Energy, 
Office of Science, Office of Nuclear Physics, under contract number DE-AC02-05CH11231.
A.B. has been supported by the Ministry of Science and Higher Education (PL), and the National Science Centre (PL), Grant No. 2023/51/B/ST2/01625.
\end{acknowledgments}
\ \\

\appendix
\section{Reduced coefficients in the presence of short-range correlations and global conservation}

To illustrate the effect of short-range correlation on reduced correlation coefficients, we consider here a system with short-range correlations with global charge conservation.
Thus, there are two sources of correlations: (i) short-range correlations, such as the thermal grand-canonical fluctuations associated with the correlation length, and (ii) long-range correlations due to global charge conservation.
For simplicity, here we consider particles only and neglect the antiparticles, which is appropriate for low collision energies.
The behavior of ordinary cumulants in such a case has been worked out in Refs.~\cite{Vovchenko:2020tsr} while factorial cumulants were analyzed in Refs.~\cite{Barej:2022jij,Barej:2022ccb}.

In the absence of global charge conservation~(``grand-canonical'' limit), the short-range correlations lead to factorial cumulants, which are linearly proportional to the mean number of particles $\mean{N}$. 
We introduce coefficients $\omega_k$ to characterize the strength of short-range correlations as
\eq{
\hat{C}_k^{\rm gce} = \omega_k \mean{N}.
}
Here $\omega_1 = 1$.
The subensemble acceptance method~(SAM) of Ref.~\cite{Vovchenko:2020tsr} allows one to incorporate global charge conservation constraints, and express the final cumulants in terms of the grand-canonical one.
Rewriting the results in terms of factorial cumulants, one obtains the following expressions for the couplings
\eq{
\hat{c}_2 & = -\frac{1}{N_{\rm tot}} + \frac{1}{N_{\rm tot}} \frac{1-\alpha}{\alpha} \omega_2, \\
\hat{c}_3 & = \frac{2}{N_{\rm tot}^2} - \frac{6}{N_{\rm tot}^2}\frac{1-\alpha}{\alpha} \omega_2 + \frac{1}{N_{\rm tot}^2} \frac{(1-\alpha)(1-2\alpha)}{\alpha^2 N_{\rm tot}^2} \omega_3, \\
\hat{c}_4 & = -\frac{6}{N_{\rm tot}^3 (1 + \omega_2)} \nonumber \\
& \quad + \frac{6(6-7\alpha)}{N_{\rm tot}^3 \alpha (1+\omega_2)} \omega_2 - \frac{12(1-\alpha)(1-4\alpha)}{N_{\rm tot}^3 \alpha^2 (1+\omega_2)} \omega_2^2 \nonumber \\
& \quad + \frac{12(1-\alpha)(1-2\alpha+(2-3\alpha)\omega_2)}{N_{\rm tot}^3 \alpha^2 (1+\omega_2)} \omega_3 \nonumber \\
& \quad - \frac{3 (1-\alpha)^2 \omega_3^2}{N_{\rm tot}^3 \alpha^2 (1+\omega_2)} \nonumber \\
& \quad + \frac{(1-\alpha)[1-3(1-\alpha)\alpha]}{N_{\rm tot}^3 \alpha^3} \omega_4.
}
Here $\alpha = \mean{N}/N_{\rm tot}$ is the fraction of the system covered by the acceptance.
One can see that for vanishing short-range correlations, $\omega_k = 0$ for $k > 1$, the couplings $\hat{c}_k$ are independent of the acceptance $\alpha$ and reduce to Eqs.~\eqref{eq:cons2}, \eqref{eq:cons3}, and \eqref{eq:cons4} obtained earlier for the ideal gas.
However, when short-range correlations are present, the couplings depend on $\alpha$, and the scaling is broken.

\bibliography{main}

\begin{thebibliography}{43}%
\makeatletter
\providecommand \@ifxundefined [1]{%
 \@ifx{#1\undefined}
}%
\providecommand \@ifnum [1]{%
 \ifnum #1\expandafter \@firstoftwo
 \else \expandafter \@secondoftwo
 \fi
}%
\providecommand \@ifx [1]{%
 \ifx #1\expandafter \@firstoftwo
 \else \expandafter \@secondoftwo
 \fi
}%
\providecommand \natexlab [1]{#1}%
\providecommand \enquote  [1]{``#1''}%
\providecommand \bibnamefont  [1]{#1}%
\providecommand \bibfnamefont [1]{#1}%
\providecommand \citenamefont [1]{#1}%
\providecommand \href@noop [0]{\@secondoftwo}%
\providecommand \href [0]{\begingroup \@sanitize@url \@href}%
\providecommand \@href[1]{\@@startlink{#1}\@@href}%
\providecommand \@@href[1]{\endgroup#1\@@endlink}%
\providecommand \@sanitize@url [0]{\catcode `\\12\catcode `\$12\catcode `\&12\catcode `\#12\catcode `\^12\catcode `\_12\catcode `\%12\relax}%
\providecommand \@@startlink[1]{}%
\providecommand \@@endlink[0]{}%
\providecommand \url  [0]{\begingroup\@sanitize@url \@url }%
\providecommand \@url [1]{\endgroup\@href {#1}{\urlprefix }}%
\providecommand \urlprefix  [0]{URL }%
\providecommand \Eprint [0]{\href }%
\providecommand \doibase [0]{http://dx.doi.org/}%
\providecommand \selectlanguage [0]{\@gobble}%
\providecommand \bibinfo  [0]{\@secondoftwo}%
\providecommand \bibfield  [0]{\@secondoftwo}%
\providecommand \translation [1]{[#1]}%
\providecommand \BibitemOpen [0]{}%
\providecommand \bibitemStop [0]{}%
\providecommand \bibitemNoStop [0]{.\EOS\space}%
\providecommand \EOS [0]{\spacefactor3000\relax}%
\providecommand \BibitemShut  [1]{\csname bibitem#1\endcsname}%
\let\auto@bib@innerbib\@empty
\bibitem [{\citenamefont {Bzdak}\ \emph {et~al.}(2020)\citenamefont {Bzdak}, \citenamefont {Esumi}, \citenamefont {Koch}, \citenamefont {Liao}, \citenamefont {Stephanov},\ and\ \citenamefont {Xu}}]{Bzdak:2019pkr}%
  \BibitemOpen
  \bibfield  {author} {\bibinfo {author} {\bibfnamefont {A.}~\bibnamefont {Bzdak}}, \bibinfo {author} {\bibfnamefont {S.}~\bibnamefont {Esumi}}, \bibinfo {author} {\bibfnamefont {V.}~\bibnamefont {Koch}}, \bibinfo {author} {\bibfnamefont {J.}~\bibnamefont {Liao}}, \bibinfo {author} {\bibfnamefont {M.}~\bibnamefont {Stephanov}}, \ and\ \bibinfo {author} {\bibfnamefont {N.}~\bibnamefont {Xu}},\ }\href {\doibase 10.1016/j.physrep.2020.01.005} {\bibfield  {journal} {\bibinfo  {journal} {Phys. Rept.}\ }\textbf {\bibinfo {volume} {853}},\ \bibinfo {pages} {1} (\bibinfo {year} {2020})},\ \Eprint {http://arxiv.org/abs/1906.00936} {arXiv:1906.00936 [nucl-th]} \BibitemShut {NoStop}%
\bibitem [{\citenamefont {Braun-Munzinger}\ \emph {et~al.}(2021)\citenamefont {Braun-Munzinger}, \citenamefont {Friman}, \citenamefont {Redlich}, \citenamefont {Rustamov},\ and\ \citenamefont {Stachel}}]{Braun-Munzinger:2020jbk}%
  \BibitemOpen
  \bibfield  {author} {\bibinfo {author} {\bibfnamefont {P.}~\bibnamefont {Braun-Munzinger}}, \bibinfo {author} {\bibfnamefont {B.}~\bibnamefont {Friman}}, \bibinfo {author} {\bibfnamefont {K.}~\bibnamefont {Redlich}}, \bibinfo {author} {\bibfnamefont {A.}~\bibnamefont {Rustamov}}, \ and\ \bibinfo {author} {\bibfnamefont {J.}~\bibnamefont {Stachel}},\ }\href {\doibase 10.1016/j.nuclphysa.2021.122141} {\bibfield  {journal} {\bibinfo  {journal} {Nucl. Phys. A}\ }\textbf {\bibinfo {volume} {1008}},\ \bibinfo {pages} {122141} (\bibinfo {year} {2021})},\ \Eprint {http://arxiv.org/abs/2007.02463} {arXiv:2007.02463 [nucl-th]} \BibitemShut {NoStop}%
\bibitem [{\citenamefont {Vovchenko}\ \emph {et~al.}(2022)\citenamefont {Vovchenko}, \citenamefont {Koch},\ and\ \citenamefont {Shen}}]{Vovchenko:2021kxx}%
  \BibitemOpen
  \bibfield  {author} {\bibinfo {author} {\bibfnamefont {V.}~\bibnamefont {Vovchenko}}, \bibinfo {author} {\bibfnamefont {V.}~\bibnamefont {Koch}}, \ and\ \bibinfo {author} {\bibfnamefont {C.}~\bibnamefont {Shen}},\ }\href {\doibase 10.1103/PhysRevC.105.014904} {\bibfield  {journal} {\bibinfo  {journal} {Phys. Rev. C}\ }\textbf {\bibinfo {volume} {105}},\ \bibinfo {pages} {014904} (\bibinfo {year} {2022})},\ \Eprint {http://arxiv.org/abs/2107.00163} {arXiv:2107.00163 [hep-ph]} \BibitemShut {NoStop}%
\bibitem [{\citenamefont {Bzdak}\ \emph {et~al.}(2017{\natexlab{a}})\citenamefont {Bzdak}, \citenamefont {Koch},\ and\ \citenamefont {Strodthoff}}]{Bzdak:2016sxg}%
  \BibitemOpen
  \bibfield  {author} {\bibinfo {author} {\bibfnamefont {A.}~\bibnamefont {Bzdak}}, \bibinfo {author} {\bibfnamefont {V.}~\bibnamefont {Koch}}, \ and\ \bibinfo {author} {\bibfnamefont {N.}~\bibnamefont {Strodthoff}},\ }\href {\doibase 10.1103/PhysRevC.95.054906} {\bibfield  {journal} {\bibinfo  {journal} {Phys. Rev. C}\ }\textbf {\bibinfo {volume} {95}},\ \bibinfo {pages} {054906} (\bibinfo {year} {2017}{\natexlab{a}})},\ \Eprint {http://arxiv.org/abs/1607.07375} {arXiv:1607.07375 [nucl-th]} \BibitemShut {NoStop}%
\bibitem [{\citenamefont {Bzdak}\ \emph {et~al.}(2017{\natexlab{b}})\citenamefont {Bzdak}, \citenamefont {Koch},\ and\ \citenamefont {Skokov}}]{Bzdak:2016jxo}%
  \BibitemOpen
  \bibfield  {author} {\bibinfo {author} {\bibfnamefont {A.}~\bibnamefont {Bzdak}}, \bibinfo {author} {\bibfnamefont {V.}~\bibnamefont {Koch}}, \ and\ \bibinfo {author} {\bibfnamefont {V.}~\bibnamefont {Skokov}},\ }\href {\doibase 10.1140/epjc/s10052-017-4847-0} {\bibfield  {journal} {\bibinfo  {journal} {Eur. Phys. J. C}\ }\textbf {\bibinfo {volume} {77}},\ \bibinfo {pages} {288} (\bibinfo {year} {2017}{\natexlab{b}})},\ \Eprint {http://arxiv.org/abs/1612.05128} {arXiv:1612.05128 [nucl-th]} \BibitemShut {NoStop}%
\bibitem [{\citenamefont {Bzdak}\ and\ \citenamefont {Koch}(2017)}]{Bzdak:2017ltv}%
  \BibitemOpen
  \bibfield  {author} {\bibinfo {author} {\bibfnamefont {A.}~\bibnamefont {Bzdak}}\ and\ \bibinfo {author} {\bibfnamefont {V.}~\bibnamefont {Koch}},\ }\href {\doibase 10.1103/PhysRevC.96.054905} {\bibfield  {journal} {\bibinfo  {journal} {Phys. Rev. C}\ }\textbf {\bibinfo {volume} {96}},\ \bibinfo {pages} {054905} (\bibinfo {year} {2017})},\ \Eprint {http://arxiv.org/abs/1707.02640} {arXiv:1707.02640 [nucl-th]} \BibitemShut {NoStop}%
\bibitem [{\citenamefont {Barej}\ and\ \citenamefont {Bzdak}(2020)}]{Barej:2020ymr}%
  \BibitemOpen
  \bibfield  {author} {\bibinfo {author} {\bibfnamefont {M.}~\bibnamefont {Barej}}\ and\ \bibinfo {author} {\bibfnamefont {A.}~\bibnamefont {Bzdak}},\ }\href {\doibase 10.1103/PhysRevC.102.064908} {\bibfield  {journal} {\bibinfo  {journal} {Phys. Rev. C}\ }\textbf {\bibinfo {volume} {102}},\ \bibinfo {pages} {064908} (\bibinfo {year} {2020})},\ \Eprint {http://arxiv.org/abs/2006.02836} {arXiv:2006.02836 [nucl-th]} \BibitemShut {NoStop}%
\bibitem [{\citenamefont {Abdallah}\ \emph {et~al.}(2021)\citenamefont {Abdallah} \emph {et~al.}}]{STAR:2021iop}%
  \BibitemOpen
  \bibfield  {author} {\bibinfo {author} {\bibfnamefont {M.}~\bibnamefont {Abdallah}} \emph {et~al.} (\bibinfo {collaboration} {STAR}),\ }\href {\doibase 10.1103/PhysRevC.104.024902} {\bibfield  {journal} {\bibinfo  {journal} {Phys. Rev. C}\ }\textbf {\bibinfo {volume} {104}},\ \bibinfo {pages} {024902} (\bibinfo {year} {2021})},\ \Eprint {http://arxiv.org/abs/2101.12413} {arXiv:2101.12413 [nucl-ex]} \BibitemShut {NoStop}%
\bibitem [{\citenamefont {Kitazawa}\ and\ \citenamefont {Asakawa}(2012{\natexlab{a}})}]{Kitazawa:2011wh}%
  \BibitemOpen
  \bibfield  {author} {\bibinfo {author} {\bibfnamefont {M.}~\bibnamefont {Kitazawa}}\ and\ \bibinfo {author} {\bibfnamefont {M.}~\bibnamefont {Asakawa}},\ }\href {\doibase 10.1103/PhysRevC.85.021901} {\bibfield  {journal} {\bibinfo  {journal} {Phys. Rev. C}\ }\textbf {\bibinfo {volume} {85}},\ \bibinfo {pages} {021901} (\bibinfo {year} {2012}{\natexlab{a}})},\ \Eprint {http://arxiv.org/abs/1107.2755} {arXiv:1107.2755 [nucl-th]} \BibitemShut {NoStop}%
\bibitem [{\citenamefont {Kitazawa}\ and\ \citenamefont {Asakawa}(2012{\natexlab{b}})}]{Kitazawa:2012at}%
  \BibitemOpen
  \bibfield  {author} {\bibinfo {author} {\bibfnamefont {M.}~\bibnamefont {Kitazawa}}\ and\ \bibinfo {author} {\bibfnamefont {M.}~\bibnamefont {Asakawa}},\ }\href {\doibase 10.1103/PhysRevC.86.024904} {\bibfield  {journal} {\bibinfo  {journal} {Phys. Rev. C}\ }\textbf {\bibinfo {volume} {86}},\ \bibinfo {pages} {024904} (\bibinfo {year} {2012}{\natexlab{b}})},\ \bibinfo {note} {[Erratum: Phys.Rev.C 86, 069902 (2012)]},\ \Eprint {http://arxiv.org/abs/1205.3292} {arXiv:1205.3292 [nucl-th]} \BibitemShut {NoStop}%
\bibitem [{\citenamefont {Bzdak}\ and\ \citenamefont {Koch}(2012)}]{Bzdak:2012ab}%
  \BibitemOpen
  \bibfield  {author} {\bibinfo {author} {\bibfnamefont {A.}~\bibnamefont {Bzdak}}\ and\ \bibinfo {author} {\bibfnamefont {V.}~\bibnamefont {Koch}},\ }\href {\doibase 10.1103/PhysRevC.86.044904} {\bibfield  {journal} {\bibinfo  {journal} {Phys. Rev. C}\ }\textbf {\bibinfo {volume} {86}},\ \bibinfo {pages} {044904} (\bibinfo {year} {2012})},\ \Eprint {http://arxiv.org/abs/1206.4286} {arXiv:1206.4286 [nucl-th]} \BibitemShut {NoStop}%
\bibitem [{\citenamefont {Savchuk}\ \emph {et~al.}(2020)\citenamefont {Savchuk}, \citenamefont {Poberezhnyuk}, \citenamefont {Vovchenko},\ and\ \citenamefont {Gorenstein}}]{Savchuk:2019xfg}%
  \BibitemOpen
  \bibfield  {author} {\bibinfo {author} {\bibfnamefont {O.}~\bibnamefont {Savchuk}}, \bibinfo {author} {\bibfnamefont {R.~V.}\ \bibnamefont {Poberezhnyuk}}, \bibinfo {author} {\bibfnamefont {V.}~\bibnamefont {Vovchenko}}, \ and\ \bibinfo {author} {\bibfnamefont {M.~I.}\ \bibnamefont {Gorenstein}},\ }\href {\doibase 10.1103/PhysRevC.101.024917} {\bibfield  {journal} {\bibinfo  {journal} {Phys. Rev. C}\ }\textbf {\bibinfo {volume} {101}},\ \bibinfo {pages} {024917} (\bibinfo {year} {2020})},\ \Eprint {http://arxiv.org/abs/1911.03426} {arXiv:1911.03426 [hep-ph]} \BibitemShut {NoStop}%
\bibitem [{\citenamefont {Pruneau}\ \emph {et~al.}(2002)\citenamefont {Pruneau}, \citenamefont {Gavin},\ and\ \citenamefont {Voloshin}}]{Pruneau:2002yf}%
  \BibitemOpen
  \bibfield  {author} {\bibinfo {author} {\bibfnamefont {C.}~\bibnamefont {Pruneau}}, \bibinfo {author} {\bibfnamefont {S.}~\bibnamefont {Gavin}}, \ and\ \bibinfo {author} {\bibfnamefont {S.}~\bibnamefont {Voloshin}},\ }\href {\doibase 10.1103/PhysRevC.66.044904} {\bibfield  {journal} {\bibinfo  {journal} {Phys. Rev. C}\ }\textbf {\bibinfo {volume} {66}},\ \bibinfo {pages} {044904} (\bibinfo {year} {2002})},\ \Eprint {http://arxiv.org/abs/nucl-ex/0204011} {arXiv:nucl-ex/0204011} \BibitemShut {NoStop}%
\bibitem [{\citenamefont {Bzdak}\ \emph {et~al.}(2013)\citenamefont {Bzdak}, \citenamefont {Koch},\ and\ \citenamefont {Skokov}}]{Bzdak:2012an}%
  \BibitemOpen
  \bibfield  {author} {\bibinfo {author} {\bibfnamefont {A.}~\bibnamefont {Bzdak}}, \bibinfo {author} {\bibfnamefont {V.}~\bibnamefont {Koch}}, \ and\ \bibinfo {author} {\bibfnamefont {V.}~\bibnamefont {Skokov}},\ }\href {\doibase 10.1103/PhysRevC.87.014901} {\bibfield  {journal} {\bibinfo  {journal} {Phys. Rev. C}\ }\textbf {\bibinfo {volume} {87}},\ \bibinfo {pages} {014901} (\bibinfo {year} {2013})},\ \Eprint {http://arxiv.org/abs/1203.4529} {arXiv:1203.4529 [hep-ph]} \BibitemShut {NoStop}%
\bibitem [{\citenamefont {Abramowitz}\ and\ \citenamefont {Stegun}(1965)}]{AbramowitzStegun}%
  \BibitemOpen
  \bibfield  {author} {\bibinfo {author} {\bibfnamefont {M.}~\bibnamefont {Abramowitz}}\ and\ \bibinfo {author} {\bibfnamefont {I.~A.}\ \bibnamefont {Stegun}},\ }in\ \href@noop {} {\emph {\bibinfo {booktitle} {US Department of Commerce}}}\ (\bibinfo  {publisher} {National Bureau of Standards Applied Mathematics series 55},\ \bibinfo {year} {1965})\BibitemShut {NoStop}%
\bibitem [{\citenamefont {Gorenstein}\ and\ \citenamefont {Gazdzicki}(2011)}]{Gorenstein:2011vq}%
  \BibitemOpen
  \bibfield  {author} {\bibinfo {author} {\bibfnamefont {M.~I.}\ \bibnamefont {Gorenstein}}\ and\ \bibinfo {author} {\bibfnamefont {M.}~\bibnamefont {Gazdzicki}},\ }\href {\doibase 10.1103/PhysRevC.84.014904} {\bibfield  {journal} {\bibinfo  {journal} {Phys. Rev. C}\ }\textbf {\bibinfo {volume} {84}},\ \bibinfo {pages} {014904} (\bibinfo {year} {2011})},\ \Eprint {http://arxiv.org/abs/1101.4865} {arXiv:1101.4865 [nucl-th]} \BibitemShut {NoStop}%
\bibitem [{\citenamefont {Holzmann}\ \emph {et~al.}(2024)\citenamefont {Holzmann}, \citenamefont {Koch}, \citenamefont {Rustamov},\ and\ \citenamefont {Stroth}}]{Holzmann:2024wyd}%
  \BibitemOpen
  \bibfield  {author} {\bibinfo {author} {\bibfnamefont {R.}~\bibnamefont {Holzmann}}, \bibinfo {author} {\bibfnamefont {V.}~\bibnamefont {Koch}}, \bibinfo {author} {\bibfnamefont {A.}~\bibnamefont {Rustamov}}, \ and\ \bibinfo {author} {\bibfnamefont {J.}~\bibnamefont {Stroth}},\ }\href {\doibase 10.1016/j.nuclphysa.2024.122924} {\bibfield  {journal} {\bibinfo  {journal} {Nucl. Phys. A}\ }\textbf {\bibinfo {volume} {1050}},\ \bibinfo {pages} {122924} (\bibinfo {year} {2024})},\ \Eprint {http://arxiv.org/abs/2403.03598} {arXiv:2403.03598 [nucl-th]} \BibitemShut {NoStop}%
\bibitem [{\citenamefont {Skokov}\ \emph {et~al.}(2013)\citenamefont {Skokov}, \citenamefont {Friman},\ and\ \citenamefont {Redlich}}]{Skokov:2012ds}%
  \BibitemOpen
  \bibfield  {author} {\bibinfo {author} {\bibfnamefont {V.}~\bibnamefont {Skokov}}, \bibinfo {author} {\bibfnamefont {B.}~\bibnamefont {Friman}}, \ and\ \bibinfo {author} {\bibfnamefont {K.}~\bibnamefont {Redlich}},\ }\href {\doibase 10.1103/PhysRevC.88.034911} {\bibfield  {journal} {\bibinfo  {journal} {Phys. Rev. C}\ }\textbf {\bibinfo {volume} {88}},\ \bibinfo {pages} {034911} (\bibinfo {year} {2013})},\ \Eprint {http://arxiv.org/abs/1205.4756} {arXiv:1205.4756 [hep-ph]} \BibitemShut {NoStop}%
\bibitem [{\citenamefont {Haiman}\ and\ \citenamefont {Schmitt}(1989)}]{haiman1989incidence}%
  \BibitemOpen
  \bibfield  {author} {\bibinfo {author} {\bibfnamefont {M.}~\bibnamefont {Haiman}}\ and\ \bibinfo {author} {\bibfnamefont {W.}~\bibnamefont {Schmitt}},\ }\href {\doibase 10.1016/0097-3165(89)90013-7} {\bibfield  {journal} {\bibinfo  {journal} {Journal of Combinatorial Theory, Series A}\ }\textbf {\bibinfo {volume} {50}},\ \bibinfo {pages} {172} (\bibinfo {year} {1989})}\BibitemShut {NoStop}%
\bibitem [{\citenamefont {Vovchenko}(2022)}]{Vovchenko:2021yen}%
  \BibitemOpen
  \bibfield  {author} {\bibinfo {author} {\bibfnamefont {V.}~\bibnamefont {Vovchenko}},\ }\href {\doibase 10.1103/PhysRevC.105.014903} {\bibfield  {journal} {\bibinfo  {journal} {Phys. Rev. C}\ }\textbf {\bibinfo {volume} {105}},\ \bibinfo {pages} {014903} (\bibinfo {year} {2022})},\ \Eprint {http://arxiv.org/abs/2106.13775} {arXiv:2106.13775 [hep-ph]} \BibitemShut {NoStop}%
\bibitem [{\citenamefont {Broniowski}\ and\ \citenamefont {Olszewski}(2017)}]{Broniowski:2017tjq}%
  \BibitemOpen
  \bibfield  {author} {\bibinfo {author} {\bibfnamefont {W.}~\bibnamefont {Broniowski}}\ and\ \bibinfo {author} {\bibfnamefont {A.}~\bibnamefont {Olszewski}},\ }\href {\doibase 10.1103/PhysRevC.95.064910} {\bibfield  {journal} {\bibinfo  {journal} {Phys. Rev. C}\ }\textbf {\bibinfo {volume} {95}},\ \bibinfo {pages} {064910} (\bibinfo {year} {2017})},\ \Eprint {http://arxiv.org/abs/1704.01532} {arXiv:1704.01532 [nucl-th]} \BibitemShut {NoStop}%
\bibitem [{\citenamefont {Sangaline}(2015)}]{Sangaline:2015bma}%
  \BibitemOpen
  \bibfield  {author} {\bibinfo {author} {\bibfnamefont {E.}~\bibnamefont {Sangaline}},\ }\href@noop {} {\  (\bibinfo {year} {2015})},\ \Eprint {http://arxiv.org/abs/1505.00261} {arXiv:1505.00261 [nucl-th]} \BibitemShut {NoStop}%
\bibitem [{\citenamefont {Adam}\ \emph {et~al.}(2021)\citenamefont {Adam} \emph {et~al.}}]{STAR:2020tga}%
  \BibitemOpen
  \bibfield  {author} {\bibinfo {author} {\bibfnamefont {J.}~\bibnamefont {Adam}} \emph {et~al.} (\bibinfo {collaboration} {STAR}),\ }\href {\doibase 10.1103/PhysRevLett.126.092301} {\bibfield  {journal} {\bibinfo  {journal} {Phys. Rev. Lett.}\ }\textbf {\bibinfo {volume} {126}},\ \bibinfo {pages} {092301} (\bibinfo {year} {2021})},\ \Eprint {http://arxiv.org/abs/2001.02852} {arXiv:2001.02852 [nucl-ex]} \BibitemShut {NoStop}%
\bibitem [{\citenamefont {Shen}\ and\ \citenamefont {Alzhrani}(2020)}]{Shen:2020jwv}%
  \BibitemOpen
  \bibfield  {author} {\bibinfo {author} {\bibfnamefont {C.}~\bibnamefont {Shen}}\ and\ \bibinfo {author} {\bibfnamefont {S.}~\bibnamefont {Alzhrani}},\ }\href {\doibase 10.1103/PhysRevC.102.014909} {\bibfield  {journal} {\bibinfo  {journal} {Phys. Rev. C}\ }\textbf {\bibinfo {volume} {102}},\ \bibinfo {pages} {014909} (\bibinfo {year} {2020})},\ \Eprint {http://arxiv.org/abs/2003.05852} {arXiv:2003.05852 [nucl-th]} \BibitemShut {NoStop}%
\bibitem [{\citenamefont {Becattini}\ \emph {et~al.}(2017)\citenamefont {Becattini}, \citenamefont {Steinheimer}, \citenamefont {Stock},\ and\ \citenamefont {Bleicher}}]{Becattini:2016xct}%
  \BibitemOpen
  \bibfield  {author} {\bibinfo {author} {\bibfnamefont {F.}~\bibnamefont {Becattini}}, \bibinfo {author} {\bibfnamefont {J.}~\bibnamefont {Steinheimer}}, \bibinfo {author} {\bibfnamefont {R.}~\bibnamefont {Stock}}, \ and\ \bibinfo {author} {\bibfnamefont {M.}~\bibnamefont {Bleicher}},\ }\href {\doibase 10.1016/j.physletb.2016.11.033} {\bibfield  {journal} {\bibinfo  {journal} {Phys. Lett. B}\ }\textbf {\bibinfo {volume} {764}},\ \bibinfo {pages} {241} (\bibinfo {year} {2017})},\ \Eprint {http://arxiv.org/abs/1605.09694} {arXiv:1605.09694 [nucl-th]} \BibitemShut {NoStop}%
\bibitem [{\citenamefont {Savchuk}\ \emph {et~al.}(2022)\citenamefont {Savchuk}, \citenamefont {Vovchenko}, \citenamefont {Koch}, \citenamefont {Steinheimer},\ and\ \citenamefont {Stoecker}}]{Savchuk:2021aog}%
  \BibitemOpen
  \bibfield  {author} {\bibinfo {author} {\bibfnamefont {O.}~\bibnamefont {Savchuk}}, \bibinfo {author} {\bibfnamefont {V.}~\bibnamefont {Vovchenko}}, \bibinfo {author} {\bibfnamefont {V.}~\bibnamefont {Koch}}, \bibinfo {author} {\bibfnamefont {J.}~\bibnamefont {Steinheimer}}, \ and\ \bibinfo {author} {\bibfnamefont {H.}~\bibnamefont {Stoecker}},\ }\href {\doibase 10.1016/j.physletb.2022.136983} {\bibfield  {journal} {\bibinfo  {journal} {Phys. Lett. B}\ }\textbf {\bibinfo {volume} {827}},\ \bibinfo {pages} {136983} (\bibinfo {year} {2022})},\ \Eprint {http://arxiv.org/abs/2106.08239} {arXiv:2106.08239 [hep-ph]} \BibitemShut {NoStop}%
\bibitem [{\citenamefont {Pruneau}(2019)}]{Pruneau:2019baa}%
  \BibitemOpen
  \bibfield  {author} {\bibinfo {author} {\bibfnamefont {C.~A.}\ \bibnamefont {Pruneau}},\ }\href {\doibase 10.1103/PhysRevC.100.034905} {\bibfield  {journal} {\bibinfo  {journal} {Phys. Rev. C}\ }\textbf {\bibinfo {volume} {100}},\ \bibinfo {pages} {034905} (\bibinfo {year} {2019})},\ \Eprint {http://arxiv.org/abs/1903.04591} {arXiv:1903.04591 [nucl-th]} \BibitemShut {NoStop}%
\bibitem [{\citenamefont {Savchuk}(2025)}]{Savchuk:2024ykb}%
  \BibitemOpen
  \bibfield  {author} {\bibinfo {author} {\bibfnamefont {O.}~\bibnamefont {Savchuk}},\ }\href {\doibase 10.1103/PhysRevC.111.024913} {\bibfield  {journal} {\bibinfo  {journal} {Phys. Rev. C}\ }\textbf {\bibinfo {volume} {111}},\ \bibinfo {pages} {024913} (\bibinfo {year} {2025})},\ \Eprint {http://arxiv.org/abs/2407.17670} {arXiv:2407.17670 [hep-ph]} \BibitemShut {NoStop}%
\bibitem [{\citenamefont {Tomasik}\ \emph {et~al.}(2019)\citenamefont {Tomasik}, \citenamefont {Melo}, \citenamefont {Laff\'ers},\ and\ \citenamefont {Bleicher}}]{Tomasik:2019ejy}%
  \BibitemOpen
  \bibfield  {author} {\bibinfo {author} {\bibfnamefont {B.}~\bibnamefont {Tomasik}}, \bibinfo {author} {\bibfnamefont {I.}~\bibnamefont {Melo}}, \bibinfo {author} {\bibfnamefont {L.}~\bibnamefont {Laff\'ers}}, \ and\ \bibinfo {author} {\bibfnamefont {M.}~\bibnamefont {Bleicher}},\ }\href {\doibase 10.22323/1.347.0155} {\bibfield  {journal} {\bibinfo  {journal} {PoS}\ }\textbf {\bibinfo {volume} {CORFU2018}},\ \bibinfo {pages} {155} (\bibinfo {year} {2019})},\ \Eprint {http://arxiv.org/abs/1903.11494} {arXiv:1903.11494 [nucl-th]} \BibitemShut {NoStop}%
\bibitem [{\citenamefont {Ling}\ and\ \citenamefont {Stephanov}(2016)}]{Ling:2015yau}%
  \BibitemOpen
  \bibfield  {author} {\bibinfo {author} {\bibfnamefont {B.}~\bibnamefont {Ling}}\ and\ \bibinfo {author} {\bibfnamefont {M.~A.}\ \bibnamefont {Stephanov}},\ }\href {\doibase 10.1103/PhysRevC.93.034915} {\bibfield  {journal} {\bibinfo  {journal} {Phys. Rev. C}\ }\textbf {\bibinfo {volume} {93}},\ \bibinfo {pages} {034915} (\bibinfo {year} {2016})},\ \Eprint {http://arxiv.org/abs/1512.09125} {arXiv:1512.09125 [nucl-th]} \BibitemShut {NoStop}%
\bibitem [{\citenamefont {Acharya}\ \emph {et~al.}(2020)\citenamefont {Acharya} \emph {et~al.}}]{ALICE:2019nbs}%
  \BibitemOpen
  \bibfield  {author} {\bibinfo {author} {\bibfnamefont {S.}~\bibnamefont {Acharya}} \emph {et~al.} (\bibinfo {collaboration} {ALICE}),\ }\href {\doibase 10.1016/j.physletb.2020.135564} {\bibfield  {journal} {\bibinfo  {journal} {Phys. Lett. B}\ }\textbf {\bibinfo {volume} {807}},\ \bibinfo {pages} {135564} (\bibinfo {year} {2020})},\ \Eprint {http://arxiv.org/abs/1910.14396} {arXiv:1910.14396 [nucl-ex]} \BibitemShut {NoStop}%
\bibitem [{\citenamefont {Acharya}\ \emph {et~al.}(2023)\citenamefont {Acharya} \emph {et~al.}}]{ALICE:2022xpf}%
  \BibitemOpen
  \bibfield  {author} {\bibinfo {author} {\bibfnamefont {S.}~\bibnamefont {Acharya}} \emph {et~al.} (\bibinfo {collaboration} {ALICE}),\ }\href {\doibase 10.1016/j.physletb.2022.137545} {\bibfield  {journal} {\bibinfo  {journal} {Phys. Lett. B}\ }\textbf {\bibinfo {volume} {844}},\ \bibinfo {pages} {137545} (\bibinfo {year} {2023})},\ \Eprint {http://arxiv.org/abs/2206.03343} {arXiv:2206.03343 [nucl-ex]} \BibitemShut {NoStop}%
\bibitem [{\citenamefont {Vovchenko}\ and\ \citenamefont {Koch}(2021)}]{Vovchenko:2020kwg}%
  \BibitemOpen
  \bibfield  {author} {\bibinfo {author} {\bibfnamefont {V.}~\bibnamefont {Vovchenko}}\ and\ \bibinfo {author} {\bibfnamefont {V.}~\bibnamefont {Koch}},\ }\href {\doibase 10.1103/PhysRevC.103.044903} {\bibfield  {journal} {\bibinfo  {journal} {Phys. Rev. C}\ }\textbf {\bibinfo {volume} {103}},\ \bibinfo {pages} {044903} (\bibinfo {year} {2021})},\ \Eprint {http://arxiv.org/abs/2012.09954} {arXiv:2012.09954 [hep-ph]} \BibitemShut {NoStop}%
\bibitem [{\citenamefont {Braun-Munzinger}\ \emph {et~al.}(2024)\citenamefont {Braun-Munzinger}, \citenamefont {Redlich}, \citenamefont {Rustamov},\ and\ \citenamefont {Stachel}}]{Braun-Munzinger:2023gsd}%
  \BibitemOpen
  \bibfield  {author} {\bibinfo {author} {\bibfnamefont {P.}~\bibnamefont {Braun-Munzinger}}, \bibinfo {author} {\bibfnamefont {K.}~\bibnamefont {Redlich}}, \bibinfo {author} {\bibfnamefont {A.}~\bibnamefont {Rustamov}}, \ and\ \bibinfo {author} {\bibfnamefont {J.}~\bibnamefont {Stachel}},\ }\href {\doibase 10.1007/JHEP08(2024)113} {\bibfield  {journal} {\bibinfo  {journal} {JHEP}\ }\textbf {\bibinfo {volume} {08}},\ \bibinfo {pages} {113} (\bibinfo {year} {2024})},\ \Eprint {http://arxiv.org/abs/2312.15534} {arXiv:2312.15534 [nucl-th]} \BibitemShut {NoStop}%
\bibitem [{\citenamefont {Vovchenko}(2024)}]{Vovchenko:2024pvk}%
  \BibitemOpen
  \bibfield  {author} {\bibinfo {author} {\bibfnamefont {V.}~\bibnamefont {Vovchenko}},\ }\href {\doibase 10.1103/PhysRevC.110.L061902} {\bibfield  {journal} {\bibinfo  {journal} {Phys. Rev. C}\ }\textbf {\bibinfo {volume} {110}},\ \bibinfo {pages} {L061902} (\bibinfo {year} {2024})},\ \Eprint {http://arxiv.org/abs/2409.01397} {arXiv:2409.01397 [hep-ph]} \BibitemShut {NoStop}%
\bibitem [{\citenamefont {Luo}\ \emph {et~al.}(2013)\citenamefont {Luo}, \citenamefont {Xu}, \citenamefont {Mohanty},\ and\ \citenamefont {Xu}}]{Luo:2013bmi}%
  \BibitemOpen
  \bibfield  {author} {\bibinfo {author} {\bibfnamefont {X.}~\bibnamefont {Luo}}, \bibinfo {author} {\bibfnamefont {J.}~\bibnamefont {Xu}}, \bibinfo {author} {\bibfnamefont {B.}~\bibnamefont {Mohanty}}, \ and\ \bibinfo {author} {\bibfnamefont {N.}~\bibnamefont {Xu}},\ }\href {\doibase 10.1088/0954-3899/40/10/105104} {\bibfield  {journal} {\bibinfo  {journal} {J. Phys. G}\ }\textbf {\bibinfo {volume} {40}},\ \bibinfo {pages} {105104} (\bibinfo {year} {2013})},\ \Eprint {http://arxiv.org/abs/1302.2332} {arXiv:1302.2332 [nucl-ex]} \BibitemShut {NoStop}%
\bibitem [{\citenamefont {Bass}\ \emph {et~al.}(2000)\citenamefont {Bass}, \citenamefont {Danielewicz},\ and\ \citenamefont {Pratt}}]{Bass:2000az}%
  \BibitemOpen
  \bibfield  {author} {\bibinfo {author} {\bibfnamefont {S.~A.}\ \bibnamefont {Bass}}, \bibinfo {author} {\bibfnamefont {P.}~\bibnamefont {Danielewicz}}, \ and\ \bibinfo {author} {\bibfnamefont {S.}~\bibnamefont {Pratt}},\ }\href {\doibase 10.1103/PhysRevLett.85.2689} {\bibfield  {journal} {\bibinfo  {journal} {Phys. Rev. Lett.}\ }\textbf {\bibinfo {volume} {85}},\ \bibinfo {pages} {2689} (\bibinfo {year} {2000})},\ \Eprint {http://arxiv.org/abs/nucl-th/0005044} {arXiv:nucl-th/0005044} \BibitemShut {NoStop}%
\bibitem [{\citenamefont {Pruneau}\ \emph {et~al.}(2023)\citenamefont {Pruneau}, \citenamefont {Gonzalez}, \citenamefont {Hanley}, \citenamefont {Marin},\ and\ \citenamefont {Basu}}]{Pruneau:2022mui}%
  \BibitemOpen
  \bibfield  {author} {\bibinfo {author} {\bibfnamefont {C.}~\bibnamefont {Pruneau}}, \bibinfo {author} {\bibfnamefont {V.}~\bibnamefont {Gonzalez}}, \bibinfo {author} {\bibfnamefont {B.}~\bibnamefont {Hanley}}, \bibinfo {author} {\bibfnamefont {A.}~\bibnamefont {Marin}}, \ and\ \bibinfo {author} {\bibfnamefont {S.}~\bibnamefont {Basu}},\ }\href {\doibase 10.1103/PhysRevC.107.014902} {\bibfield  {journal} {\bibinfo  {journal} {Phys. Rev. C}\ }\textbf {\bibinfo {volume} {107}},\ \bibinfo {pages} {014902} (\bibinfo {year} {2023})},\ \Eprint {http://arxiv.org/abs/2209.10420} {arXiv:2209.10420 [hep-ph]} \BibitemShut {NoStop}%
\bibitem [{\citenamefont {Kuznietsov}\ \emph {et~al.}(2022)\citenamefont {Kuznietsov}, \citenamefont {Savchuk}, \citenamefont {Gorenstein}, \citenamefont {Koch},\ and\ \citenamefont {Vovchenko}}]{Kuznietsov:2022pcn}%
  \BibitemOpen
  \bibfield  {author} {\bibinfo {author} {\bibfnamefont {V.~A.}\ \bibnamefont {Kuznietsov}}, \bibinfo {author} {\bibfnamefont {O.}~\bibnamefont {Savchuk}}, \bibinfo {author} {\bibfnamefont {M.~I.}\ \bibnamefont {Gorenstein}}, \bibinfo {author} {\bibfnamefont {V.}~\bibnamefont {Koch}}, \ and\ \bibinfo {author} {\bibfnamefont {V.}~\bibnamefont {Vovchenko}},\ }\href {\doibase 10.1103/PhysRevC.105.044903} {\bibfield  {journal} {\bibinfo  {journal} {Phys. Rev. C}\ }\textbf {\bibinfo {volume} {105}},\ \bibinfo {pages} {044903} (\bibinfo {year} {2022})},\ \Eprint {http://arxiv.org/abs/2201.08486} {arXiv:2201.08486 [hep-ph]} \BibitemShut {NoStop}%
\bibitem [{\citenamefont {Kuznietsov}\ \emph {et~al.}(2024)\citenamefont {Kuznietsov}, \citenamefont {Gorenstein}, \citenamefont {Koch},\ and\ \citenamefont {Vovchenko}}]{Kuznietsov:2024xyn}%
  \BibitemOpen
  \bibfield  {author} {\bibinfo {author} {\bibfnamefont {V.~A.}\ \bibnamefont {Kuznietsov}}, \bibinfo {author} {\bibfnamefont {M.~I.}\ \bibnamefont {Gorenstein}}, \bibinfo {author} {\bibfnamefont {V.}~\bibnamefont {Koch}}, \ and\ \bibinfo {author} {\bibfnamefont {V.}~\bibnamefont {Vovchenko}},\ }\href {\doibase 10.1103/PhysRevC.110.015206} {\bibfield  {journal} {\bibinfo  {journal} {Phys. Rev. C}\ }\textbf {\bibinfo {volume} {110}},\ \bibinfo {pages} {015206} (\bibinfo {year} {2024})},\ \Eprint {http://arxiv.org/abs/2404.00476} {arXiv:2404.00476 [nucl-th]} \BibitemShut {NoStop}%
\bibitem [{\citenamefont {Vovchenko}\ \emph {et~al.}(2020)\citenamefont {Vovchenko}, \citenamefont {Savchuk}, \citenamefont {Poberezhnyuk}, \citenamefont {Gorenstein},\ and\ \citenamefont {Koch}}]{Vovchenko:2020tsr}%
  \BibitemOpen
  \bibfield  {author} {\bibinfo {author} {\bibfnamefont {V.}~\bibnamefont {Vovchenko}}, \bibinfo {author} {\bibfnamefont {O.}~\bibnamefont {Savchuk}}, \bibinfo {author} {\bibfnamefont {R.~V.}\ \bibnamefont {Poberezhnyuk}}, \bibinfo {author} {\bibfnamefont {M.~I.}\ \bibnamefont {Gorenstein}}, \ and\ \bibinfo {author} {\bibfnamefont {V.}~\bibnamefont {Koch}},\ }\href {\doibase 10.1016/j.physletb.2020.135868} {\bibfield  {journal} {\bibinfo  {journal} {Phys. Lett. B}\ }\textbf {\bibinfo {volume} {811}},\ \bibinfo {pages} {135868} (\bibinfo {year} {2020})},\ \Eprint {http://arxiv.org/abs/2003.13905} {arXiv:2003.13905 [hep-ph]} \BibitemShut {NoStop}%
\bibitem [{\citenamefont {Barej}\ and\ \citenamefont {Bzdak}(2022)}]{Barej:2022jij}%
  \BibitemOpen
  \bibfield  {author} {\bibinfo {author} {\bibfnamefont {M.}~\bibnamefont {Barej}}\ and\ \bibinfo {author} {\bibfnamefont {A.}~\bibnamefont {Bzdak}},\ }\href {\doibase 10.1103/PhysRevC.106.024904} {\bibfield  {journal} {\bibinfo  {journal} {Phys. Rev. C}\ }\textbf {\bibinfo {volume} {106}},\ \bibinfo {pages} {024904} (\bibinfo {year} {2022})},\ \Eprint {http://arxiv.org/abs/2205.05497} {arXiv:2205.05497 [hep-ph]} \BibitemShut {NoStop}%
\bibitem [{\citenamefont {Barej}\ and\ \citenamefont {Bzdak}(2023)}]{Barej:2022ccb}%
  \BibitemOpen
  \bibfield  {author} {\bibinfo {author} {\bibfnamefont {M.}~\bibnamefont {Barej}}\ and\ \bibinfo {author} {\bibfnamefont {A.}~\bibnamefont {Bzdak}},\ }\href {\doibase 10.1103/PhysRevC.107.034914} {\bibfield  {journal} {\bibinfo  {journal} {Phys. Rev. C}\ }\textbf {\bibinfo {volume} {107}},\ \bibinfo {pages} {034914} (\bibinfo {year} {2023})},\ \Eprint {http://arxiv.org/abs/2210.15394} {arXiv:2210.15394 [hep-ph]} \BibitemShut {NoStop}%
\end{thebibliography}%

\end{document}